\pdfoutput=1
\documentclass{article}
\usepackage[utf8]{inputenc}
\usepackage{amsfonts}
\usepackage[mathscr]{euscript}
\usepackage{xcolor}

\usepackage[numbers]{natbib} % has a nice set of citation styles and commands
\usepackage{mathtools} % amsmath with fixes and additions
\usepackage{bbm}
\usepackage{booktabs} % commands to create good-looking tables
\usepackage{tikz} % nice language for creating drawings and diagrams
\usepackage{boxedminipage}
\usepackage{alltt}
\usepackage{bm}
\usepackage{geometry}
\usepackage{hyperref}
\usepackage{listings}
\usepackage{subcaption}
\usepackage{wrapfig}
\usepackage{authblk}

\usepackage{amsmath,amsfonts,bm}

\def\eqref#1{equation~\ref{#1}}
\def\1{\bm{1}}

\def\vg{{\bm{g}}}

\DeclareMathAlphabet{\mathsfit}{\encodingdefault}{\sfdefault}{m}{sl}
\SetMathAlphabet{\mathsfit}{bold}{\encodingdefault}{\sfdefault}{bx}{n}

\DeclareMathOperator{\defeq}{\stackrel{\text{def}}{=}}

\title{Disentangling Fact from Grid Cell Fiction\\in Trained Deep Path Integrators}

\author[1]{Rylan Schaeffer}
\author[2]{Mikail Khona}
\author[1]{Sanmi Koyejo}
\author[3,4]{Ila Rani Fiete}

\affil[1]{Computer Science, Stanford}
\affil[2]{Physics, MIT}
\affil[3]{Brain and Cognitive Sciences, MIT}
\affil[4]{McGovern Institute for Brain Research at MIT}

\date{}

\begin{document}

\maketitle

\begin{abstract}
Work on deep learning-based models of grid cells  suggests that grid cells generically and robustly arise from optimizing networks to path integrate, i.e., track one's spatial position by integrating self-velocity signals. In previous work \cite{schaeffer_no_2022}, we challenged this path integration hypothesis by showing that deep neural networks trained to path integrate almost always do so, but almost never learn grid-like tuning unless separately inserted by researchers via mechanisms unrelated to path integration. In this work, we restate the key evidence substantiating these insights, then address a response to \cite{schaeffer_no_2022} by authors of one of the path integration hypothesis papers \cite{sorscher_when_2022}. First, we show that the response misinterprets our work, indirectly confirming our points. Second, we evaluate the response's preferred ``unified theory for the origin of grid cells" in trained deep path integrators \cite{sorscher_unied_2019,sorscher_unified_2020,sorscher_unified_2022} and show that it is at best ``occasionally suggestive," not exact or comprehensive. 
We finish by considering why assessing model quality through prediction of biological neural activity by regression of activity in deep networks \cite{nayebi_explaining_2021} can lead to the wrong conclusions.
\end{abstract}

\section{Introduction: Grid Cells \& the Path Integration Hypothesis}

The discovery of grid cells \cite{hafting2005microstructure,stensola2012entorhinal} in the mammalian brain sparked considerable work on their mechanisms and origins. One approach to understanding grid cells is through the lens of optimization in deep neural networks. A number of works in this direction published in high-profile venues assert that grid cells generically and robustly emerge when recurrent networks are trained to path integrate (Nature \cite{banino_vector-based_2018}, Neuron \cite{sorscher_unified_2022}) and machine learning conferences (Neural Information Processing Systems Spotlights \cite{sorscher_unied_2019, nayebi_explaining_2021}, International Conference on Learning Representations \cite{cueva_emergence_2018}): 

\begin{itemize}
    \item Main text of \cite{banino_vector-based_2018}: “Notably, therefore, our results show that grid-like representations reminiscent of those found in the mammalian entorhinal cortex emerge in a generic network trained to path integrate.”
    \item Abstract of \cite{sorscher_unified_2020}: “Here we forge an intimate link between the computational problem of path-integration and the existence of hexagonal grids, by demonstrating that such grids arise generically in biologically plausible neural networks trained to path integrate. Moreover, we develop a unifying theory for why hexagonal grids are so ubiquitous in path-integrator circuits.”
    \item Highlights of \cite{sorscher_unified_2022}: “RNNs trained to path integrate with nonnegative firing develop hexagonal grid cells."
    \item Abstract of \cite{cueva_emergence_2018}: ``We trained recurrent neural networks (RNNs) to perform navigation tasks in 2D arenas based on velocity inputs. Surprisingly, we find that grid-like spatial response patterns emerge in trained networks."
    \item Key figure in \cite{sorscher_unified_2022}: ``Neural networks trained on normative tasks develop grid-like ﬁring ﬁelds.''
\end{itemize}

\begin{figure*}[t!]
    \centering
    \includegraphics[width=0.95\textwidth]{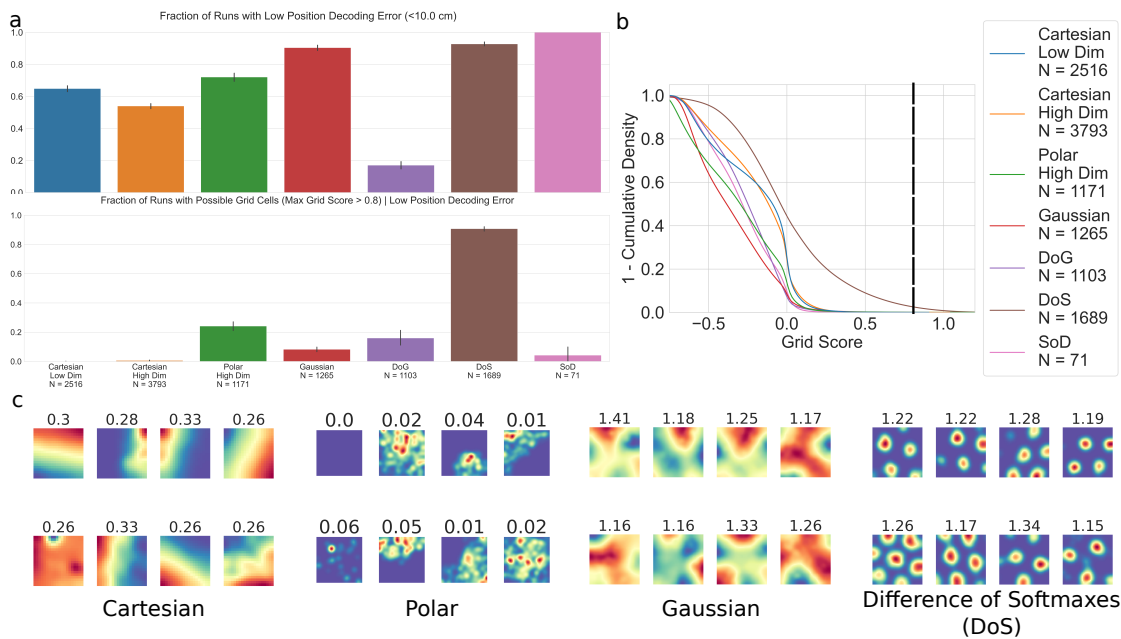}
    \caption{(a) Top: Most networks learn to path integrate, regardless of supervised target. Bottom: Few networks display possible grid-like tuning. (b) Survival functions of grid scores per supervised target. (c) Rate maps of top grid-scoring units in trained deep path integrators with i) Cartesian, ii) Polar, iii) Gaussian, iv) specifically selected (tuned) Difference-of-Softmaxes (DoS) targets.  Only DoS supervised targets induce grid cells. Numbers above rate maps are $60^{\circ}$ grid scores. Figure from \cite{schaeffer_no_2022}.}
    \label{fig:path_integration_and_tuning}
\end{figure*}

These publications collectively paint the picture that simply learning to path integrate (possibly with a non-negativity constraint) surprisingly, generically, and robustly causes the networks to learn grid-like representations matching those found in the mammalian entorhinal cortex. We call this claim the \textit{path integration hypothesis}.
However, there are reasons to question this path integration hypothesis: the hypothesis is inconsistent with prior numerical results training deep recurrent neural networks to perform spatial tasks \cite{kanitscheider_training_2016,kanitscheider_emergence_2017} and contradicts theoretical work suggesting the grid code is special not just for path integration but also for its coding-theoretic properties \cite{fiete_what_2008, sreenivasan_grid_2011, mathis_optimal_2012, wei2015principle}.

Motivated by these reasons, we rigorously and systematically investigated the path integration hypothesis in our paper ``No Free Lunch from Deep Learning in Neuroscience: A Case Study through Models of the Entorhinal-Hippocampal Circuit'' (NFL, for short) \cite{schaeffer_no_2022}. NFL combined large-scale numerical experiments and mathematical guidance to show that deep recurrent neural networks trained to path integrate frequently learn to do so, but almost never learn grid-like tuning, unless grid-like tuning is inserted via handcrafted supervised targets designed specifically to produce grid-like tuning. In this paper, we:
\begin{enumerate}
    \item Restate the key evidence from NFL substantiating these conclusions.
    \item Address a response to NFL by leading authors of the path integration hypothesis \cite{sorscher_when_2022} (henceforth, ``the Response''), explaining how the Response misinterprets NFL's findings and in doing so validates NFL's core findings.
    \item Evaluate the validity of results from the``unified theory for the origin of grid cells" \cite{sorscher_unied_2019} and demonstrate that the Unified Theory is \textit{at best} an occasionally suggestive description of trained deep path integrators.
\end{enumerate}

\section{Evidence Challenging the Path Integration Hypothesis}

\begin{figure}[t!]
    \centering
    \includegraphics[width=0.95\textwidth]{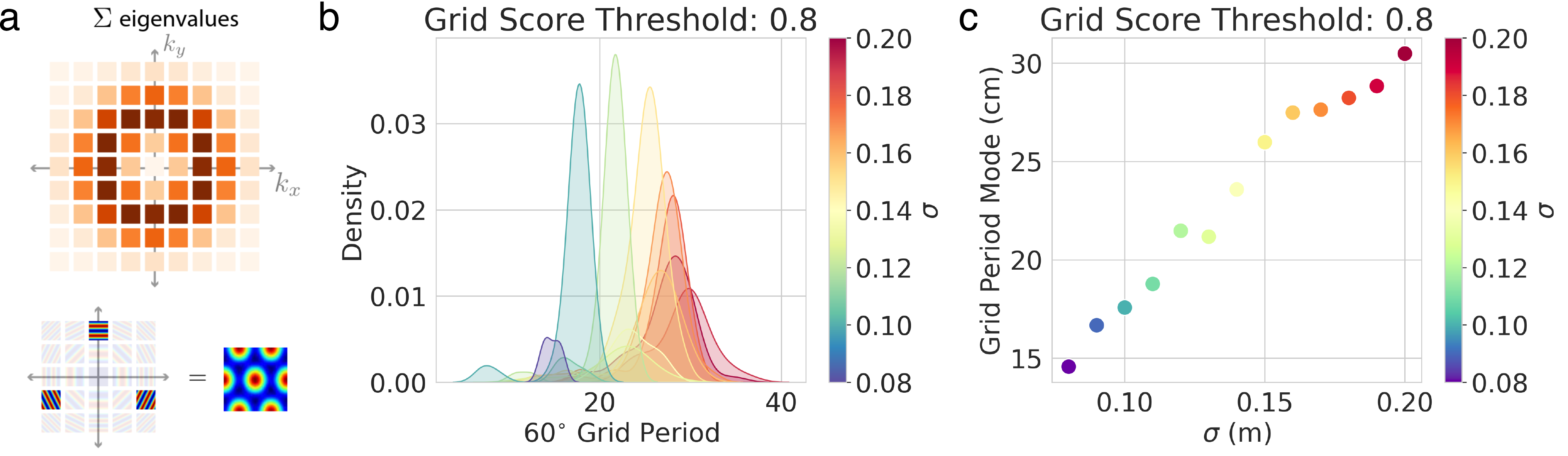}
    \caption{(a) The ``Difference of Softmaxes" (DoS) supervised target is specifically chosen to place Fourier power on an annulus of a desired radius; doing so embeds grid-like tuning in the supervised targets that deep path integrators are then trained to predict. (b) The researchers' choice of DoS targets inserts a single grid periodicity into the networks. (c) The choice of DoS hyperparameters sets the periodicity of the grid-like tuning. Subfigure (a) from \cite{sorscher_unified_2020}, (b-c) from \cite{schaeffer_no_2022}.}
    \label{fig:grid_modules}
\end{figure}

NFL challenged the path integration hypothesis by demonstrating that grid-like tuning does not emerge in deep recurrent networks trained to path integrate unless researchers insert grid-like tuning via supervised targets designed to produce grid-like tuning \cite{schaeffer_no_2022}. 

To substantiate these claims, NFL used large scale hyperparameter sweeps to train over 11,000 deep recurrent neural networks to path integrate. Most learned to successfully path integrate, but few learned \textit{possible} grid-like tuning (Fig. \ref{fig:path_integration_and_tuning}a). Grid-like tuning appeared only when researchers chose a specific and otherwise hard-to-justify supervised target called ``Difference-of-Softmaxes'' (DoS) (Fig. \ref{fig:path_integration_and_tuning}ab) that was designed to produce grid-like tuning \cite{sorscher_unied_2019, sorscher_unified_2020} (Fig. \ref{fig:grid_modules}a). This DoS target does not improve path integration performance: thus it is not a functionally motivated choice. The DoS is also not a biologically motivated choice \cite{schaeffer_testing_2023}. NFL demonstrated that the choice of DoS inserts into the networks a single grid periodicity (Fig. \ref{fig:grid_modules}b), and the grid periodicity is directly set by the researchers' chosen DoS hyperparameters (Fig. \ref{fig:grid_modules}c).
Further, \cite{sorscher_unied_2019} notes that multiple grid modules, a key property of biological grid cells \cite{stensola2012entorhinal}, do not emerge from their framework, which NFL confirmed, meaning one cannot evaluate whether the artificial grid modules follow biological grid module ratios.

Based on this and additional evidence, NFL concluded that grid-like tuning in trained deep path integrators is often a post-hoc result in which losses, target functions, and regularizers are explored until the result resembles grid cells. 

\section{The Response to NFL \cite{sorscher_when_2022} Misinterprets Then Corroborates NFL \cite{schaeffer_no_2022}}

The message of multiple prominent papers is that grid cells naturally result from a simple path integration task \cite{sorscher_unied_2019, sorscher_unified_2020,nayebi_explaining_2021, sorscher_unified_2022}; these papers underemphasize or omit that \textit{the} key ingredient for grid-like tuning in deep networks is researchers designing supervised targets to produce grid-like tuning. Even after the publication of NFL, a Neuron paper \cite{sorscher_unified_2022} neglected mentioning the critical importance of the DoG or DoS supervised targets in the Highlights, Summary and Introduction while emphasizing the less-central claim about path integration: ``Here, we forge a link between the problem of path integration and the existence of hexagonal grids, by demonstrating that such grids arise in neural networks trained to path integrate under simple biologically plausible constraints. Moreover, we develop a unifying theory for why hexagonal grids are ubiquitous in path-integrator circuits."

The Response to NFL \cite{sorscher_when_2022} misinterprets NFL from its outset: ``[NFL] presented a sequence of simulation results and some theoretical analysis suggesting prior work involved non-transparent fine-tuning."
NFL is not about fine-tuning.
NFL is about prior work selecting improbable readout functions to drive grid-like tuning, but presenting the results as a tale of grid cells resulting naturally and directly from optimization for path integration.

The Response then corroborates NFL's central findings at length (Sec. 2 in \cite{sorscher_when_2022}).
We summarize the ``unified theory'' for completeness (App. \ref{app:theory}) and highlight our key takeaway here: \textit{The ``unified theory'' is not a theory of path integration, nor of deep recurrent neural networks, nor of learning dynamics of deep recurrent neural networks trained to path integrate. Rather, it is a theory of Fourier structure in supervised targets.}
The ``unified theory'' starts with the static function approximation model of prior work \cite{dordek_extracting_2016}, and studies learning representations $\vg$ to predict/reconstruct a given supervised target $P$:
\begin{equation}
    \min_{\vg} ||P - \vg \vg^T P||^2 + \sigma(\vg) \quad \text{ subject to } \vg^T \vg = 1.
\end{equation}

This optimization problem's minima are set by the choice of supervised targets $P$, specifically through the autocorrelation matrix $\Sigma_{xx'} = (P P^T)_{xx'}$.
The Unified Theory designs $P$ using DoS tuning curves to embed grid-like tuning into $\Sigma$'s Fourier structure (Fig. \ref{fig:grid_modules}). These supervised targets drive networks (recurrent or feedforward) to develop grid-like tuning without regard to path integration. Indeed, the authors showed shallow nonlinear autoencoders can develop grid-like tuning when trained on these targets \cite{sorscher_unied_2019}, despite the complete absence of a velocity signal.

\begin{figure}
    \centering
    \includegraphics[width=0.95\textwidth]{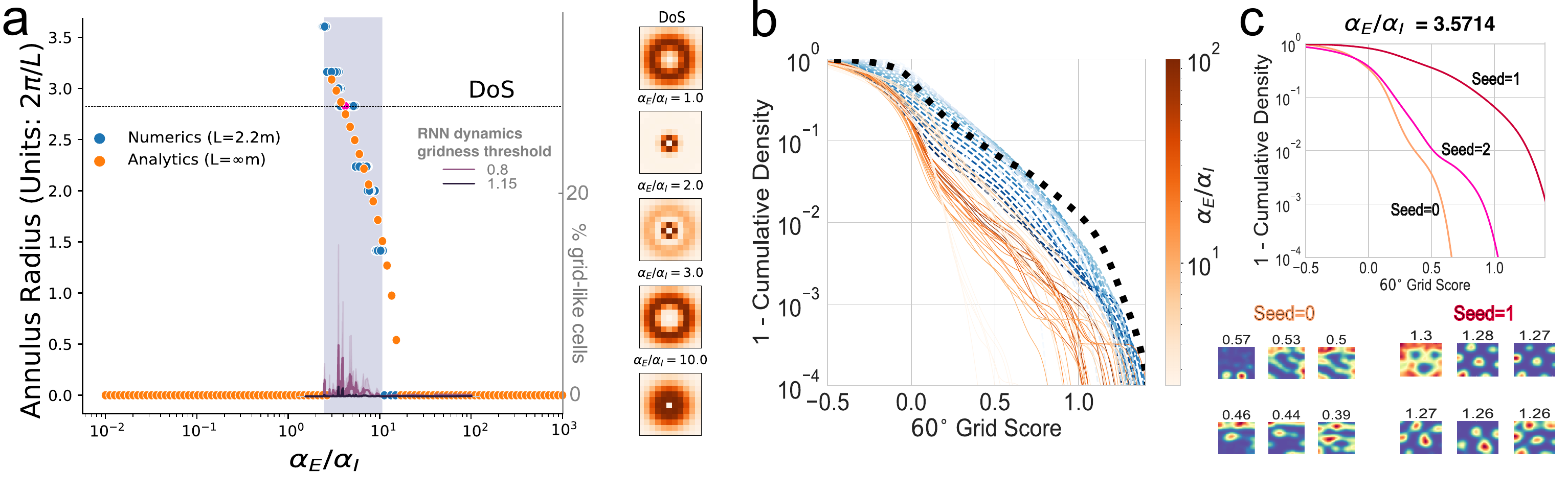}
    \caption{(a) In the Unified Theory, only a narrow range of hyperparameters $\sigma_E, \sigma_I$ should (analytically and numerically) contain Fourier power on an annulus of an appropriate radius to create grid-like tuning (right). Networks trained on Difference-of-Gaussians (DoG) supervised targets do not develop grid-like tuning outside a narrow hyperparameter interval and inconsistently develop grid-like tuning inside the hyperparameter interval. (b) Across $\alpha_E/ \alpha_I$ DoG ratios, DoG networks (orange) generally score worse than filtered-and-thresholded noise (blue) and worse than DoS (black). (c) More densely sweeping within the theoretically feasible $\alpha_E/\alpha_I$ DoG region, and choosing $\alpha_E$ and $\alpha_I$ magnitudes closely matching DoS, shown in (c) still showed only one ratio that did at least as well as a DoS. This result was true for only one out of three seeds: two ``sibling" runs with otherwise identical settings produced poor gridness. All networks learn to path integrate. Subfigures from \cite{schaeffer_no_2022}.}
    \label{fig:fourier_power_necessary_but_insufficient}
\end{figure}
% XXX is the % grid cells on the right axis referring to % of cells in each trained network that have grid-like tuning? or % of trained networks with any grid-like tuning? 
% Rylan 2023/11/21: Percent of cells in each trained network. The "dips" are because networks with hyperparameters in between didn't learn any grid cells. 
% XXX 20% grid cells in each run is arguably close to the biology, where <20% cells in MEC layer II are grid cells. It would be a better critique if 20% of runs resulted in a >10% fraction of grid cells. What is the % runs? 
% Rylan 2023/11/24: (a) The point of this figure is that the in-between values are 0. Within the interval, you don't get consistent results. (b) The % percentage of grid-like units in the interval is generously 5%, probably less.

\section{Evidence Challenging the Unified Theory \cite{sorscher_unied_2019}}

\begin{figure}[t!]
    \centering
    \includegraphics[width=0.95\textwidth]{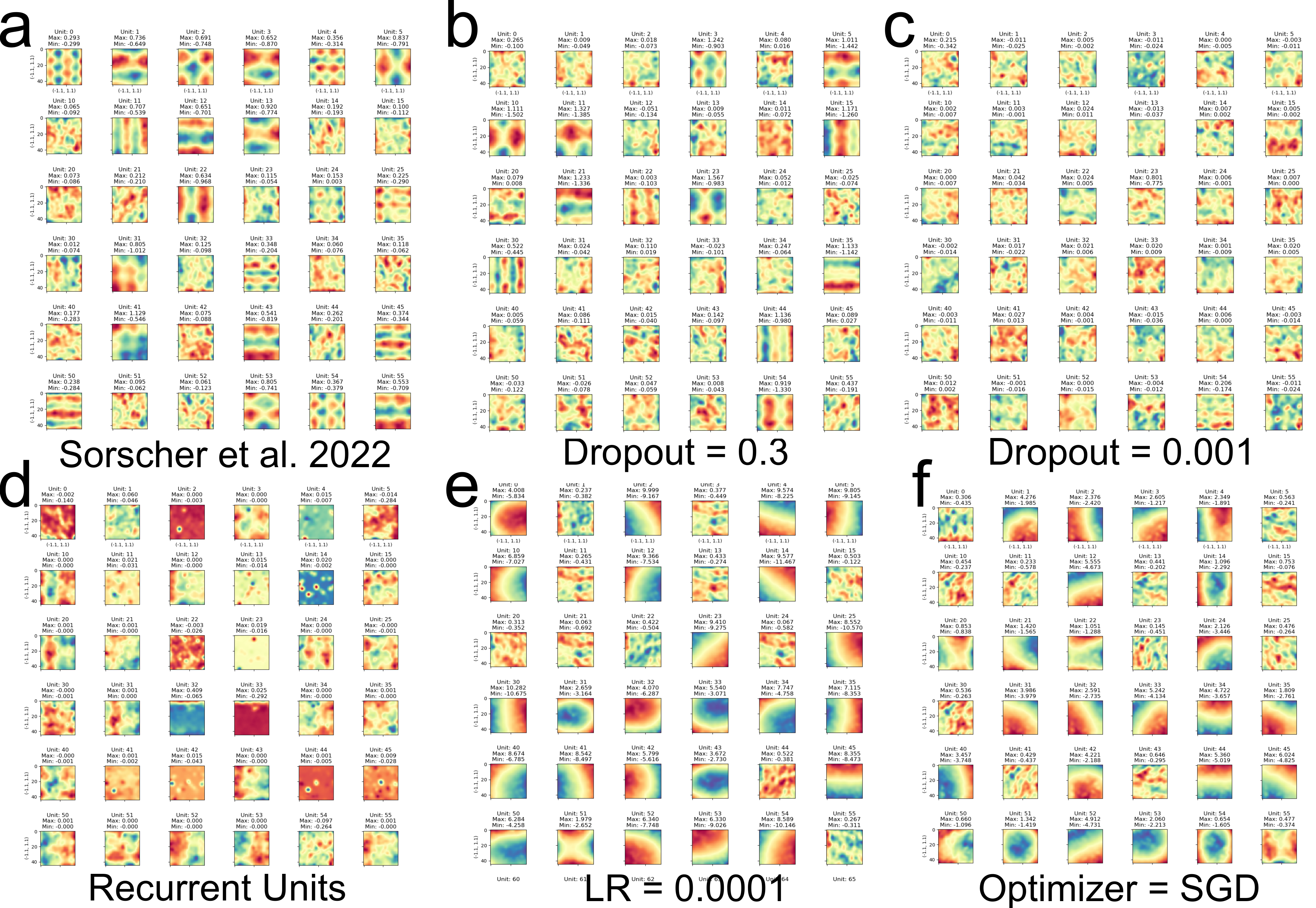}
    \caption{(a) Code by Ganguli et al. \cite{sorscher_unified_2022} to show that Gaussian targets can give (square) lattices uses a more-elaborate architecture with several additional components adapted from previous work \cite{banino_vector-based_2018},  rather the architecture of their earlier publications\cite{sorscher_unied_2019, sorscher_unified_2020, nayebi_explaining_2021}.
    (b-c) Reducing or removing dropout from this model causes lattice-like responses to disappear without affecting path integration performance. (d) The recurrent units of the architecture also display no lattices. (e-f) Changing other implementation details, such as the learning rate or optimizer, also causes the lattices to disappear without affecting path integration performance. See also Figs. \ref{app:fig:sorscher_2022_rebuttal_dropout}, \ref{app:fig:sorscher_2022_rebuttal_learning_rate}, \ref{app:fig:sorscher_2022_rebuttal_optimizer}}
    \label{fig:sorscher_refutation}
\end{figure}

The Response to NFL claimed that the Unified Theory \cite{sorscher_unied_2019}, as a complete theory for the emergence of grid cells in deep path integrators, already contained the results shown in NFL -- this is a point we disagree with. To see why, we turn to the Unified Theory. First, as noted in the previous section, the Unified Theory of grid cell emergence in deep path integrators is not that -- it is a theory of the Fourier structure of supervised targets. Second, two critical assumptions made by the Unified Theory (i.e. place cells are  translationally invariant as a population, and place cells have center-surround tuning individually) were recently shown to be biologically unlikely \cite{schaeffer_testing_2023}. Here, we set aside these two issues and explore the results of the Unified Theory and their relevance to deep path integrators, which we find to be at most ``occasionally suggestive," not exact or comprehensive.  

\paragraph{Fourier structure in supervised targets is necessary but insufficient.} The Unified Theory predicts that supervised targets with Fourier power on an annulus of an appropriate radius should result in grid-like tuning in path integrating deep networks. We derive the spectrum for Difference-of-Gaussians (DoG) targets and show that only a narrow slice of hyperparameters possess the appropriate radius (Fig. \ref{fig:fourier_power_necessary_but_insufficient}a). But even when networks are trained inside this narrow interval, they only occasionally develop grid-like tuning (Fig. \ref{fig:fourier_power_necessary_but_insufficient}ab).
% Deep networks trained outside this narrow hyperparameter interval achieve equal path integration performance but do not develop grid-like tuning.
The very specific Difference-of-Softmaxes (DoS) leads to grid-like tuning more often than DoG target (Fig. \ref{fig:fourier_power_necessary_but_insufficient}b), but grid-like tuning still does not reliably emerge and is seed dependent (Fig. \ref{fig:fourier_power_necessary_but_insufficient}c). These findings show that the Unified Theory's conditions are not sufficient. To reiterate, the criticism is not that deep learning requires hyperparameter tuning, but rather that a sequence of highly-specific and biologically-implausible choices orthogonal to path integration make clear that grid cell tuning was the goal of the entire setup, and that grid cells did not simply pop out as a natural consequence of path integration. 
%  was implicitly treated as a target on which the hyperparameters were optimized, and thus they were 

\paragraph{Gaussian-like supervised targets do not produce grid-like tuning without additional elements.}

A key point of contention is whether Gaussian supervised targets produce grid-like tuning. 
Early work claimed yes \cite{banino_vector-based_2018,sorscher_unied_2019,sorscher_unified_2020}, but NFL showed that Gaussian supervised targets almost never produce lattices (either hexagonal or square) \cite{schaeffer_no_2022}. This finding of NFL was contested by \cite{sorscher_when_2022}, which claimed that with specific hyperparameters Gaussian targets should lead to grid cells in deep path integrators, and released new code in support. Here we present three new results showing that Gaussian supervised targets in deep path integrators do not on their own produce lattices. This debate matters for two reasons. Firstly, Gaussian supervised targets might be easier to justify biologically than DoG and DoS targets, and secondly, whether Gaussian targets produce grids sheds light on the size of the space of grid-producing supervised targets.

Firstly, we deconstruct the new code released by the authors of \cite{sorscher_unified_2022}, which produces (square) lattices from Gaussian targets. This code uses a different architecture than the deep recurrent networks used in their prior work \cite{sorscher_unied_2019, sorscher_unified_2020, nayebi_explaining_2021} and is similar to the model of \cite{banino_vector-based_2018}. It is described by the authors as: ``a significantly more complex LSTM architecture [...] The ``grid cells'' in this architecture are not recurrently connected, but reside in a dense layer immediately following the LSTM. Importantly, the grid cells are now subject to dropout at a rate of 0.5.'' We found that the high dropout rate, together with specific choices of high learning rates and optimizers, are pivotal to achieving lattice-like responses when the target functions are Gaussian (Fig. \ref{fig:sorscher_refutation}, App. \ref{app:sec:sorscher_2022_rebuttal_gaussian}, Figs. \ref{app:fig:sorscher_2022_rebuttal_dropout}, \ref{app:fig:sorscher_2022_rebuttal_learning_rate}, \ref{app:fig:sorscher_2022_rebuttal_optimizer}), even though good path integration can be achieved without these features. This also bolsters NFL's critique of the narrative of grid cell emergence in \cite{banino_vector-based_2018}. 

\begin{figure}[t!]
    \centering\includegraphics[width=0.475\textwidth]{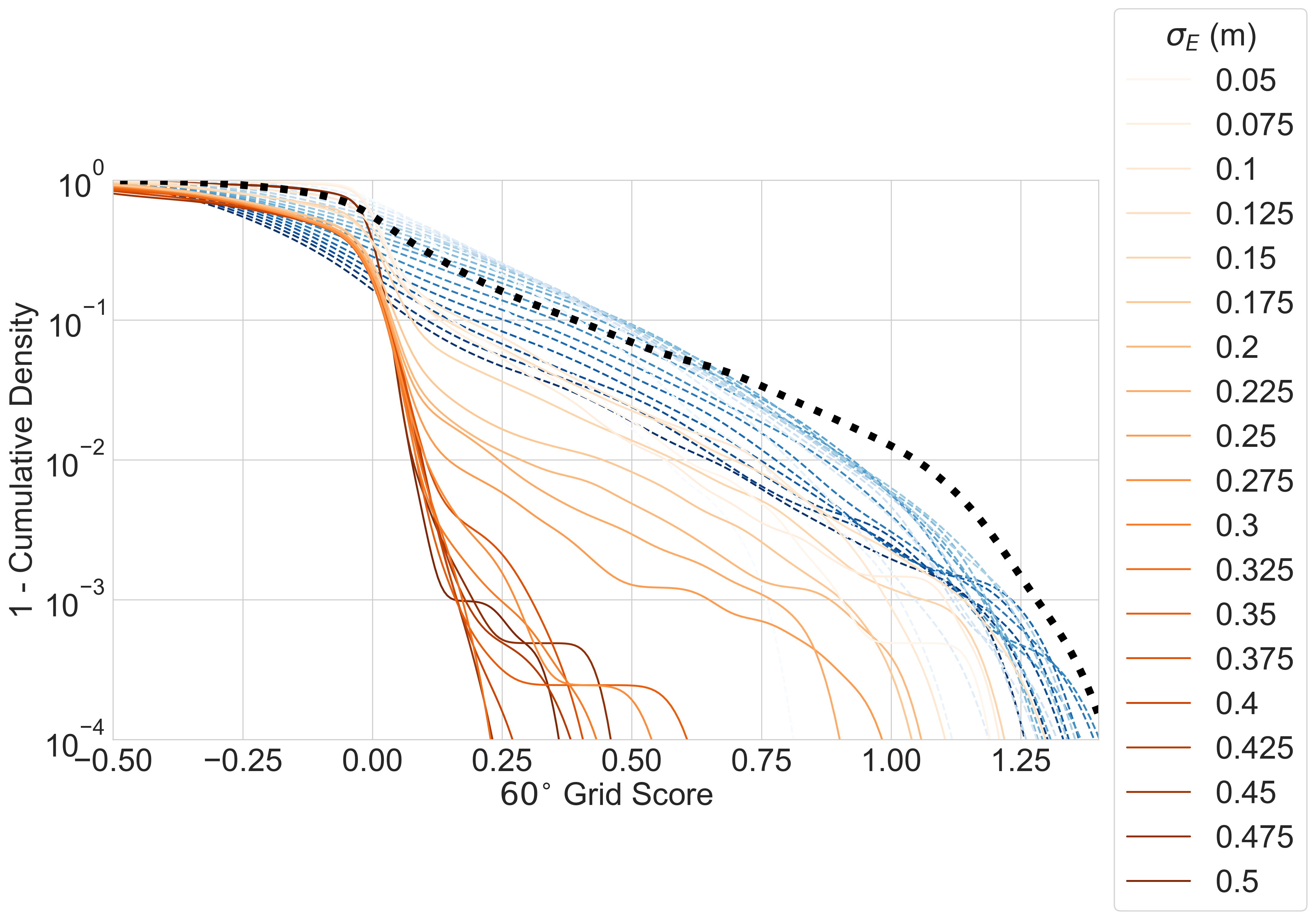}%
    \centering\includegraphics[width=0.475\textwidth]{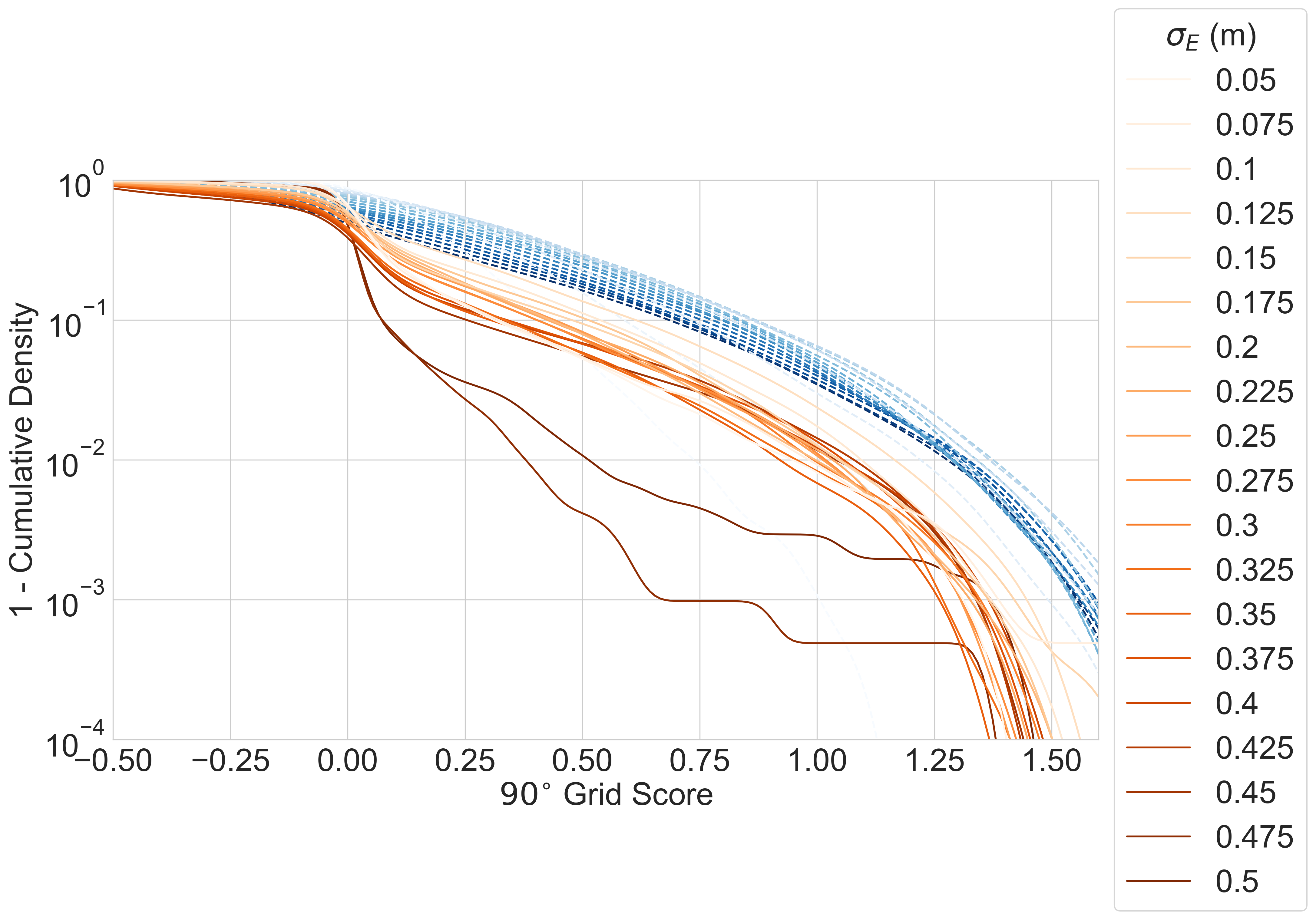}
    \caption{Deep path integrators with ReLU nonlinearities and Gaussian readouts, sweeping Gaussian scale, do not learn lattices: RNNs (Red) Gaussian readout scale was swept $\sigma_E \in [5, 50]$ cm in increments of 2.5 cm. For negative control, we consider $\text{Uniform}(-1, 1)$ noise filtered-then-ReLU-thresholded (Blue), and for positive control, we consider RNNs with ReLU nonlinearities trained on ideal-width ($12$ cm) DoS readouts (Black). RNN+ReLU+Gaussian $60^{\circ}$ score distributions are dominated by almost all noise filter widths, and the ideal-width DoS $60^{\circ}$ score distribution dominates most noise filter widths (especially at the tail). All other hyperparameters match \cite{sorscher_unified_2022}'s Page e2.
    }
    \label{fig:rnn_relu_gaussian}
\end{figure}

\begin{figure}
    \centering\includegraphics[width=0.475\textwidth]{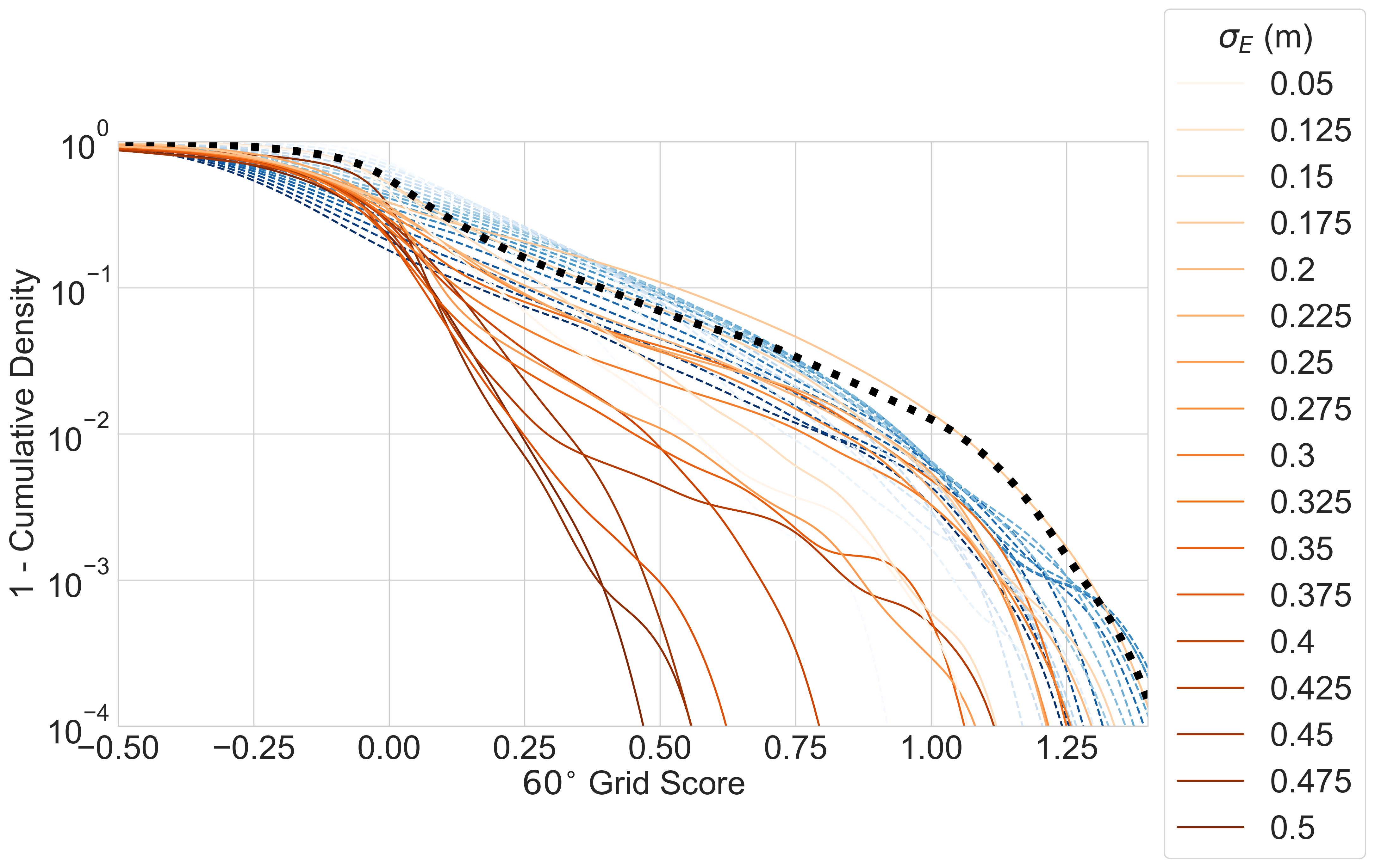}%
    \centering\includegraphics[width=0.475\textwidth]{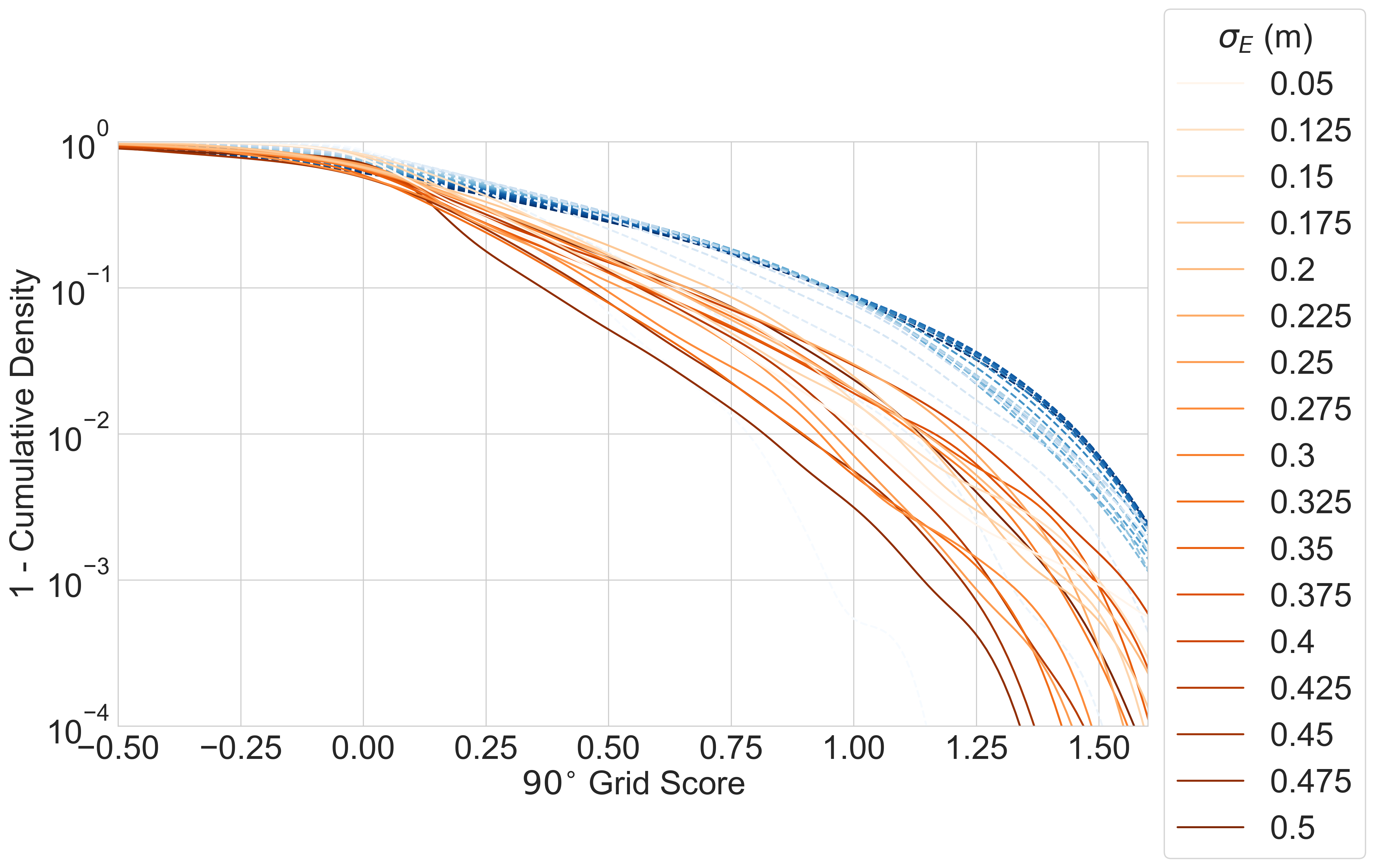}
    \caption{Deep path integrators with Gaussian readouts and Tanh nonlinearity achieve lower grid score distributions than low-pass-filtered-then-ReLU-thresholded noise: RNNs' (Red) Gaussian readout scale was swept $\sigma_E \in [5, 50]$ cm in increments of 2.5 cm. For negative control, we consider $\text{Uniform}(0, 1)$ noise filtered(Blue), and for positive control, we consider RNNs with ReLU nonlinearities trained on ideal-width ($12$ cm) DoS readouts. RNN+ReLU+Gaussian $90^{\circ}$ score distributions are dominated by almost all noise filter widths, and the ideal-width DoS $90^{\circ}$ score distribution dominates most noise filter widths (especially at the tail). All other hyperparameters match \cite{sorscher_unified_2022}'s Table on Page e2.
    }
    \label{fig:rnn_tanh_gaussian}
\end{figure}

\begin{figure}
        \centering
        \includegraphics[width=\textwidth]{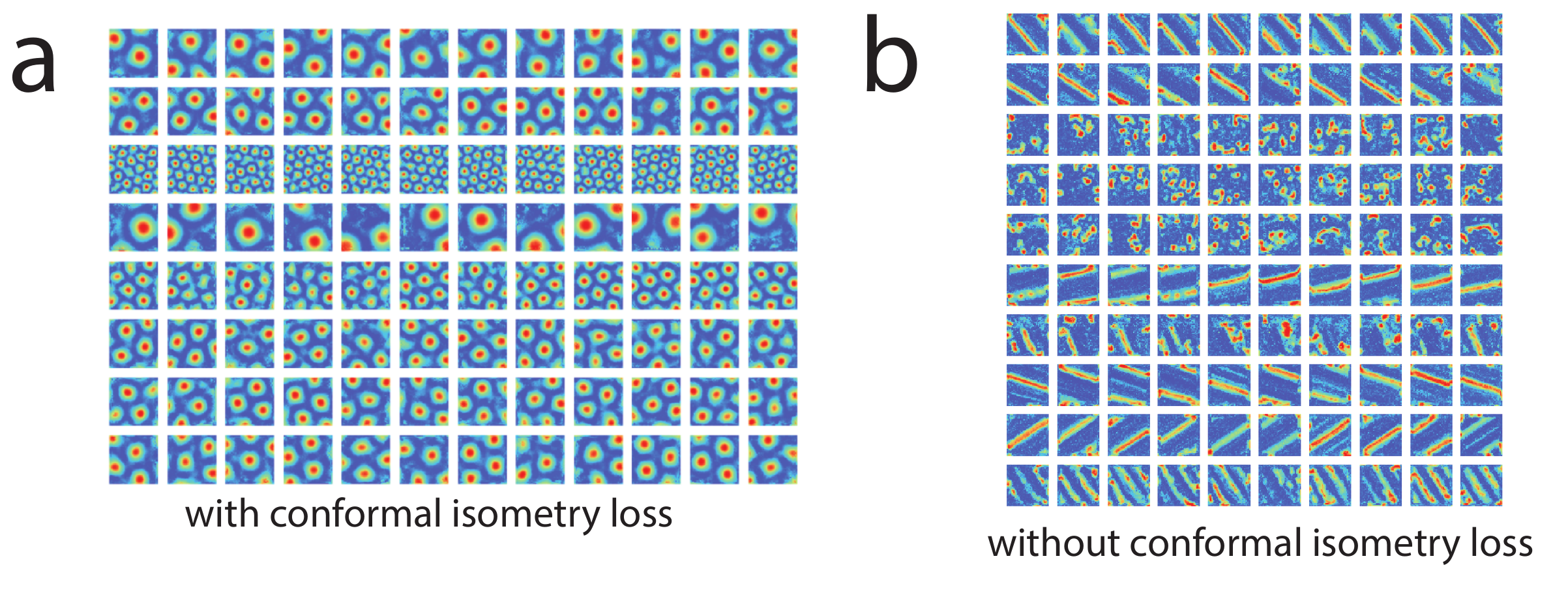}
        \caption{Independent research by Xu et al. 2022 \cite{xu2022conformal} shows Gaussian supervised targets do not create grid-like tuning \cite{xu2022conformal}. In order to obtain lattices with Gaussians, Xu et al. 2022 proposed a novel conformal isometry loss (left). Networks trained without the loss, i.e., solely with Gaussian supervised targets do not learn lattice-like units (right). Figure included with permission from authors.}
        \label{fig:xu_2022}
\end{figure}

Secondly, returning to the authors' original code from \cite{nayebi_explaining_2021}, we densely sweep the recommended Gaussian receptive field, testing (a) recurrent networks with ReLU nonlinearities that the Unified Theory predicts should learn hexagonal lattices and (b) Tanh nonlinearities that the Unified Theory predicts should learn square lattices. Following \cite{sorscher_unied_2019}, we used filtered and thresholded noise as the null distribution to determine when observed patterns were more significant than chance in all cases. Across nonlinearities and receptive field widths, the sweeps show networks do not learn grid-like tuning from Gaussian targets (Figs \ref{fig:rnn_relu_gaussian}, \ref{fig:rnn_tanh_gaussian}).

Thirdly, we highlight recent independent research by Xu et al. 2022 \cite{xu2022conformal} confirming that Gaussian supervised targets do not produce grid-like tuning (Fig. \ref{fig:xu_2022}). 

\begin{figure}
    \centering
    \includegraphics[width=0.37\textwidth]{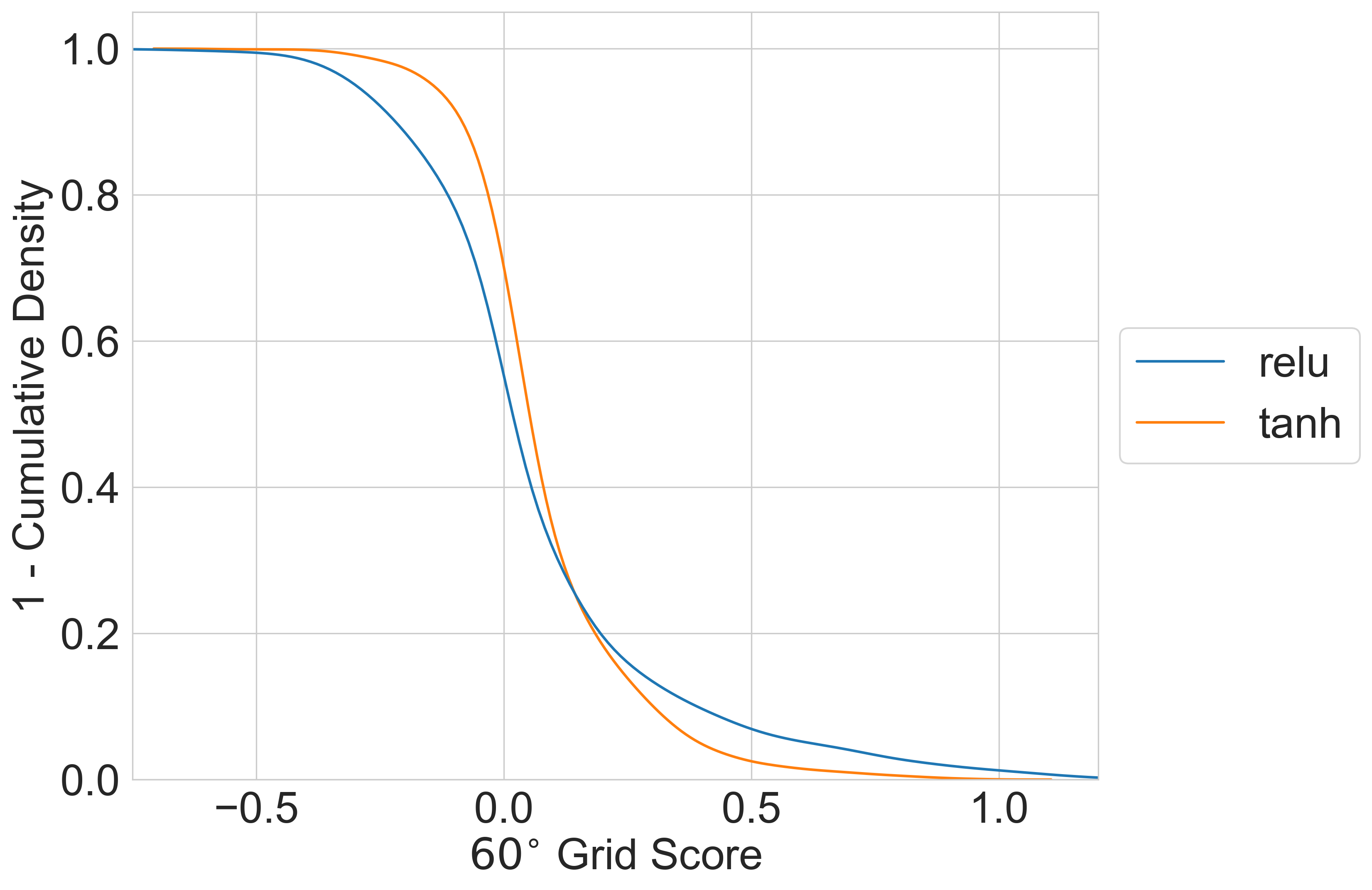}%
    \includegraphics[width=0.37\textwidth]{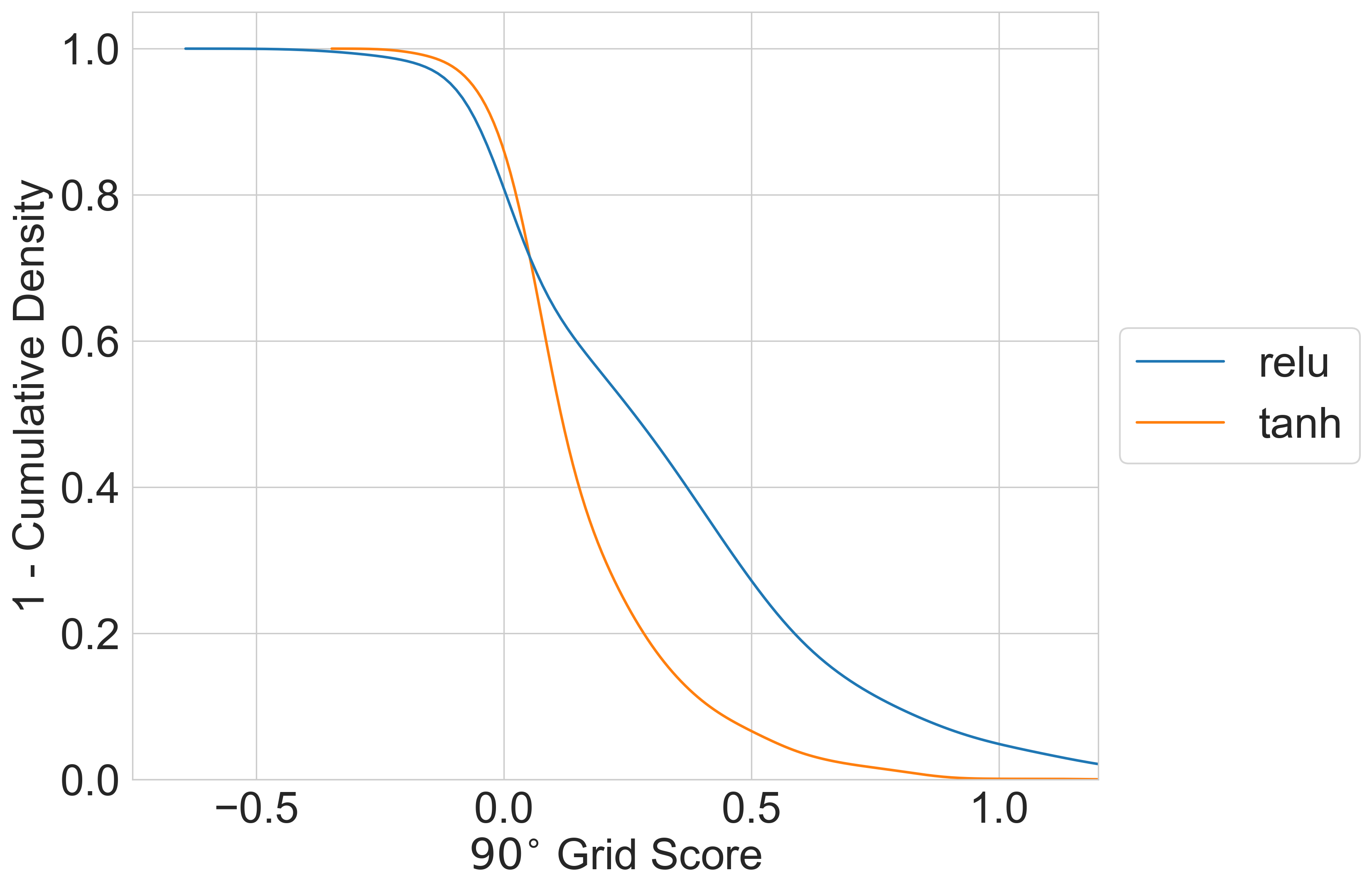}%
    \includegraphics[width=0.22\textwidth]{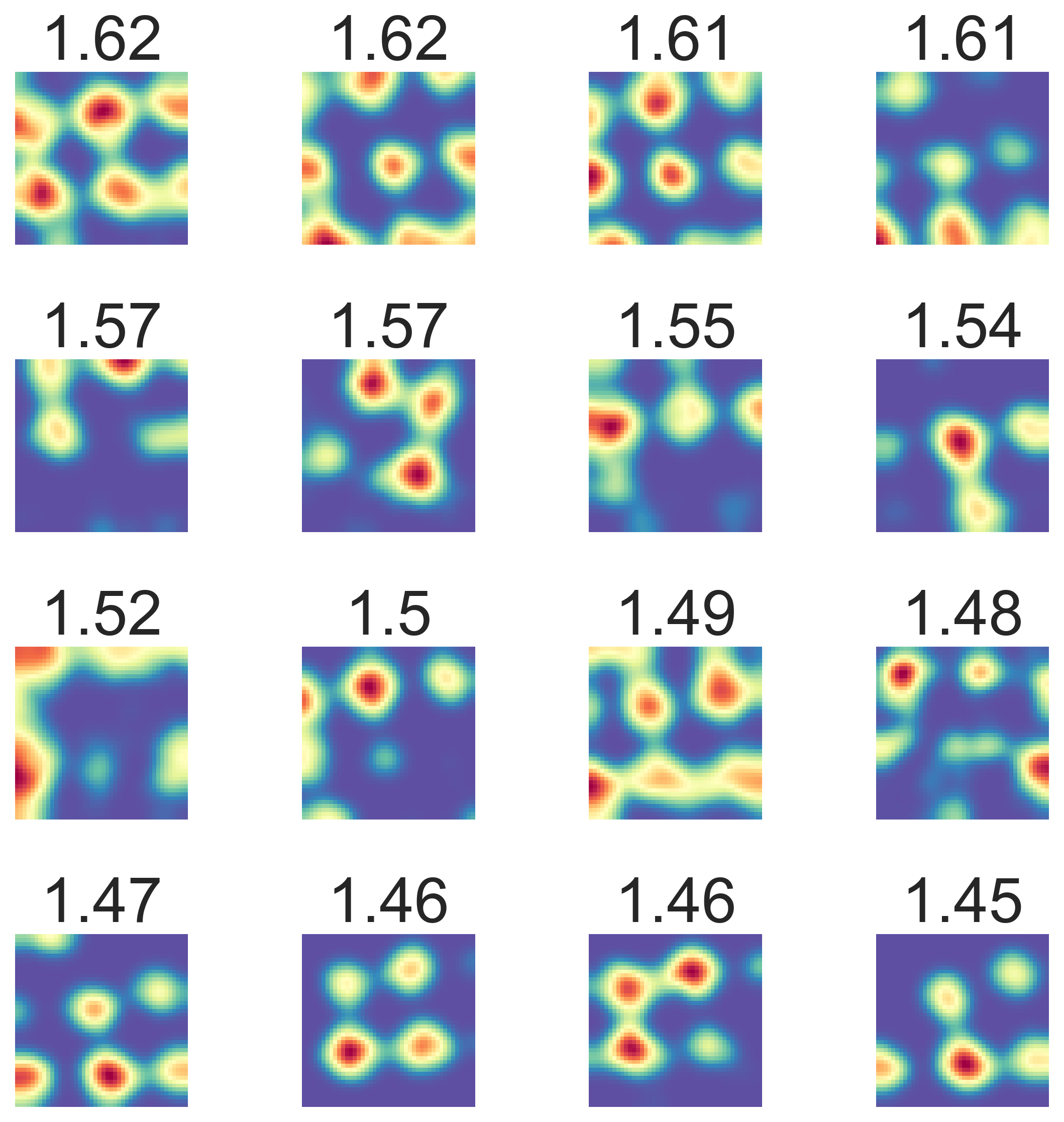}%
    \caption{Using DoS supervised targets with ideal hyperparameters, switching from Tanh to ReLU increases $60^{\circ}$ scores (left), but also increases $90^{\circ}$ scores (center). Example square rate maps (right). 3 seeds $\in \{0, 1, 2\}$ per nonlinearity; all other hyperparameters matched \cite{sorscher_unied_2019,sorscher_unified_2022}'s code defaults.}
    \label{fig:lattice_geometry}
\end{figure}

\paragraph{Contrary to Unified Theory, non-negativity produces hexagonal \textit{and} square lattices.}
A key result of the Unified Theory \cite{sorscher_unified_2022, sorscher_unied_2019} is that non-negativity favors hexagonal lattices over square lattices: ``Once a nonlinear constraint such as non-negativity is added, the optimization favors a single type of map corresponding to hexagonal grid cells."
We find that the default hexagonal grid-forming hyperparameters consistently produce networks that simultaneously learn both hexagonal \textit{and square} lattices (Fig. \ref{fig:lattice_geometry}): Replacing Tanh with ReLU for non-negativity causes an upward shift in hexagonal $60^{\circ}$ scores, but also causes a larger upward shift in square $90^{\circ}$ scores \textit{in the same networks}. Visual examination of units with high $90^{\circ}$ scores in ReLU networks confirms square tuning curves. Training here and in NFL \cite{schaeffer_no_2022} is 20$\times$ longer than in \cite{sorscher_unified_2022}, so insufficient training cannot explain the result.

\begin{figure}
    \centering
    \includegraphics[width=\textwidth]{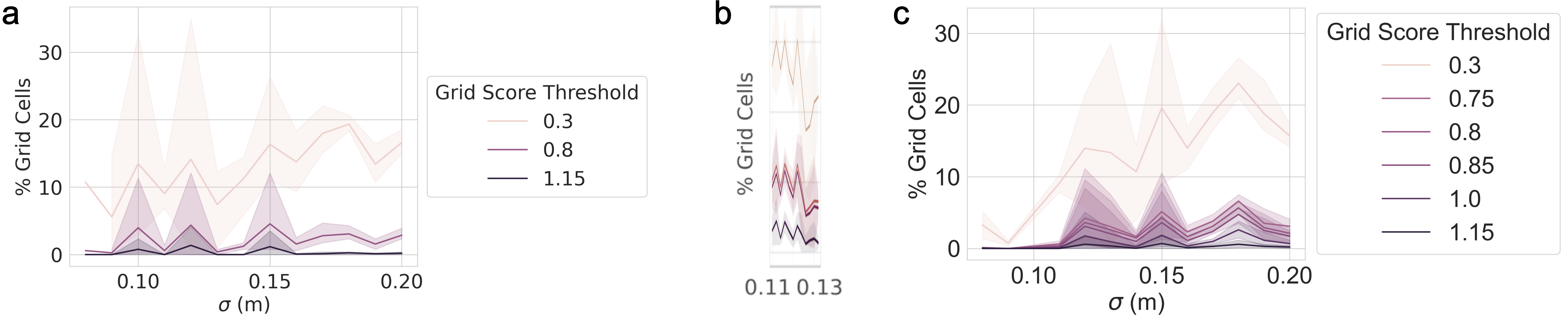}
    \caption{(a) NFL discovered small alterations to the DoS target width $\sigma_E$ can result in the disappearance of grid units \cite{schaeffer_no_2022}. (b) The Response disputes this result \cite{sorscher_when_2022}, albeit over a much narrower range of $\sigma_E$ values. (c) Rerunning our previous sweep but with larger networks (4096 units, previously 1024 units) to exactly match newly published hyperparameters \cite{sorscher_unified_2022} again shows that grid-like tuning is highly sensitive to hyperparameters. All networks achieve low position decoding error.}
    \label{fig:dos_sensitivity_to_sigmaE}
\end{figure}

\paragraph{Sensitivity to supervised target hyperparameters.}
NFL reported that even with DoS supervised targets, the emergence of grid-like tuning is highly sensitive to the receptive field width $\sigma_E$ (Fig. \ref{fig:dos_sensitivity_to_sigmaE}A) when sweeping over the range $8$cm to $20$cm \cite{schaeffer_no_2022}. The Response stated that DoS readouts are robust to $\sigma_E$ \cite{sorscher_when_2022} but showed only a narrow range of $11$cm to $13$cm (Fig. \ref{fig:dos_sensitivity_to_sigmaE}B). 
% and provided no explanation for what NFL's publicly available code does incorrectly, despite NFL's code being forked from their code in \cite{nayebi_explaining_2021} and the networks trained with matching hyperparameters, varying only $\sigma_E$.
The only noted difference in published hyperparameters between the NFL result, which was based on code from \cite{nayebi_explaining_2021}, and the Response was the number of hidden units (ours: 1024 units; their Response: 4096 units). We find that the NFL result on sensitivity to supervised target field width holds in 4096-unit networks (Fig. \ref{fig:dos_sensitivity_to_sigmaE}C).

\section{Why Can Neural Regressions Be So Confidently Wrong?}

A widely used method to test models in modern computational neuroscience is neural regression \cite{yamins2014performance, khaligh2014deep}: predicting biological neural recordings from activity in trained deep networks, typically using linear regression. Nayebi et al. 2021 \cite{nayebi_explaining_2021} used this technique to accurately predict tetrode recordings from mouse medial entorhinal cortex (MEC) from the path integrating networks and thus to conclude that their networks were excellent models of MEC. 

\begin{figure}
    \centering
    \begin{minipage}[c]{0.3\textwidth}
        \centering
        \includegraphics[width=\linewidth]{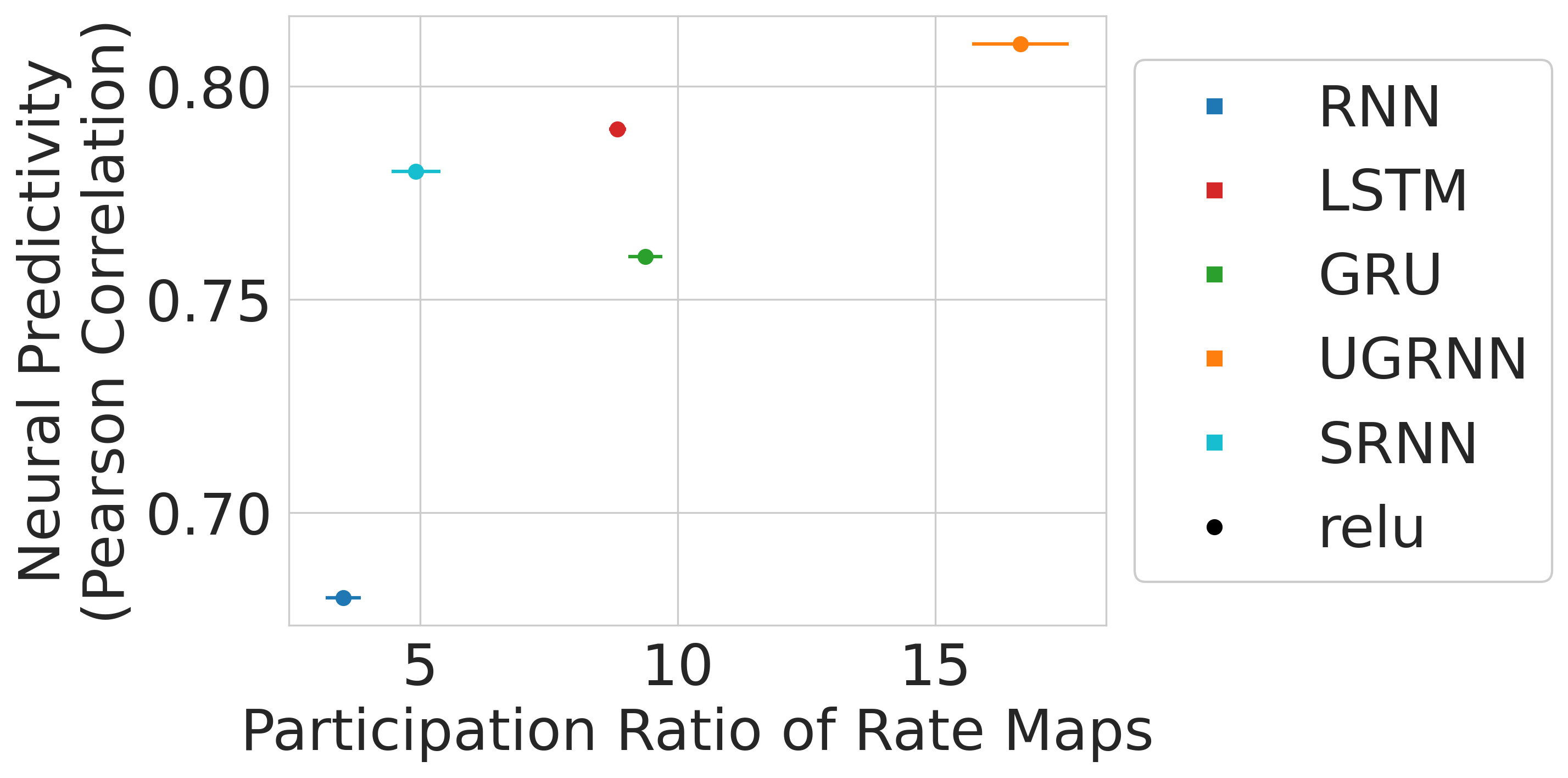}
        \label{fig:schaeffer2022}
    \end{minipage}%
    \begin{minipage}[c]{0.3\textwidth}
        \centering
        \includegraphics[width=\linewidth]{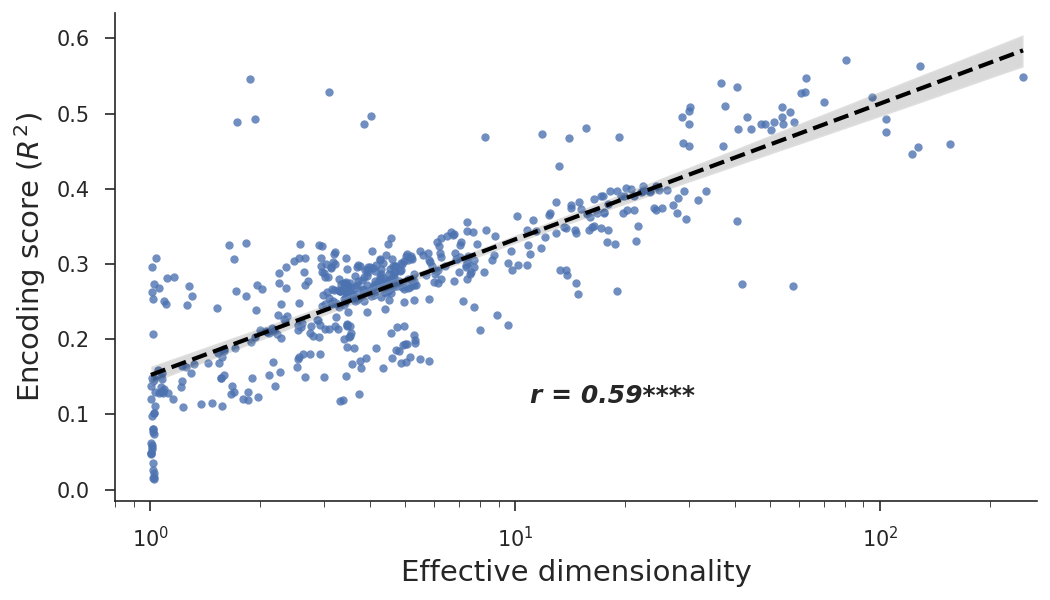}
        \label{fig:elmoznino2022}
    \end{minipage}%
    \begin{minipage}[c]{0.4\textwidth}
        \centering
        \includegraphics[width=\linewidth]{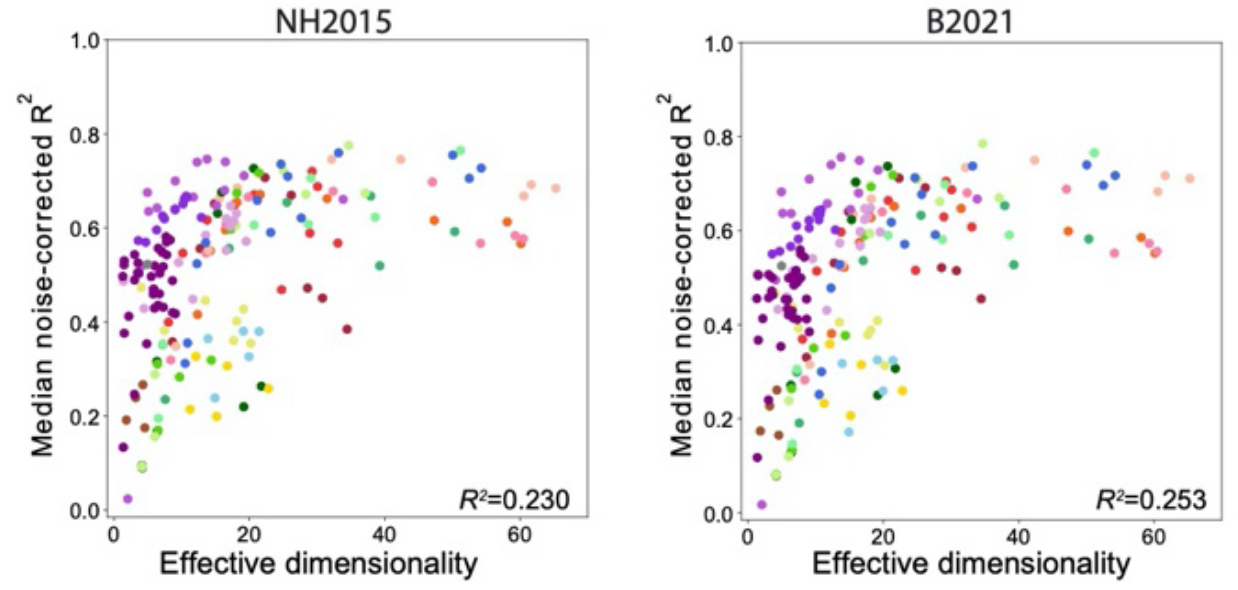}
        \label{fig:tuckute2022}
    \end{minipage}
    \caption{Three independent research labs studying three different brain circuits in three different species across three different modalities found a consistent relationship: test $R^2$ in neural regressions is approximately an affine function of the log of the effective dimensionality (Eqn. \ref{eq:effective_dimensionality}) of the artificial neural activations. Figures from (a) Spatial Navigation in Mouse Medial Entorhinal Cortex \cite{schaeffer_no_2022}, (b) Vision in Macaque IT Cortex \cite{elmoznino_high-performing_2022} (note the log-X scaling), and (c) Audition in Human fMRI \cite{tuckute_many_2022}.}
    \label{fig:findings}
\end{figure}

NFL hypothesized that trained deep networks might achieve higher neural predictivity scores than competitor models simply because trained deep networks provide higher dimensional, richer basis functions that could be able to generate closer fits to the data (and by extension better within-sample predictions of the data) than lower dimensional, poorer basis functions supplied by alternative models \cite{schaeffer_no_2022}. NFL found initial supporting evidence: the reported test Pearson correlation (termed ``neural predictivity") approximately followed a linear-log relationship with the effective dimensionality (sometimes termed intrinsic dimensionality or participation ratio) \cite{litwin-kumar_optimal_2017,recanatesi2019dimensionality,recanatesi2022scale} of the networks' representations (Fig. \ref{fig:findings}a), defined as:
\begin{equation}
    \text{Participation Ratio a.k.a. Effective Dimensionality} \quad \defeq \quad \frac{(\sum_r \lambda_r)^2}{\sum_r \lambda_r^2}.
    \label{eq:effective_dimensionality}
\end{equation}

NFL then conjectured similar results would be found in other modalities. Two independent research groups studying two different modalities (vision and audition) in two different brain circuits (visual ventral stream and auditory cortex) across two different species (macaque and human) using two different recording technologies (ECoG, fMRI) found a quantitatively similar relationship: in vision, Elmoznino et al. 2022 \cite{elmoznino_high-performing_2022} found the same approximate linear-log relationship in regressions between convolutional networks and macaque IT Cortex (Fig \ref{fig:findings}b), and in audition, Tuckute et al. 2022 \cite{tuckute_many_2022} found the same linear-log relationship regressions between deep networks and human fMRI responses (Fig \ref{fig:findings}c).

NFL and these results call the neural regression methodology into question. Han, Poggio \& Cheung 2023 \cite{han2023system} sought to validate the neural regression methodology by asking how well the methodology can identify the correct network if the correct network is included within the set of candidate models, and concluded that the results were poor. Canatar et al. 2023 \cite{canatar2023spectral} provided a theory showing candidate networks can achieve high predictivity scores so long as candidate networks have certain spectral properties relative to the target networks, without any assumption of a mechanistic, biological or otherwise meaningful relationship between the candidate and target networks. One possibility is that neural regressions are spiritually similar to Ramsey Theory \cite{katz2018introduction}: networks trained with more parameters, data and compute are more likely to offer richer basis functions such that there will exist a subset of basis functions able to predict whatever neural recordings, and we know that both underparameterized and overparameterized linear regression can generalize \cite{schaeffer2023double,schaeffer2023divergence}. Mathematically formalizing and proving such a possibility for neural regressions is non-trivial, but collectively, these results call for further investigation into what exactly the neural regression methodology tells us about neural representations.

\section{Discussion and Future Directions}

NFL, the Response, and the current document together show that path integration is an insufficient condition for the emergence of grid cells in deep neural networks. Contrary to the claims made in the various papers on grid cell emergence in deep networks, successfully training networks to path integrate does not automatically lead to grid cells -- unrelated and special ingredients must be added to obtain grid cells. Thus, to obtain grid cells in such networks requires effectively baking them in, with the researcher implicitly using grid cells as a major part of their own loss function in their investigations. \textit{While there is no harm in explicitly taking such an approach, and indeed such approaches can be enlightening when the conditions of emergence are carefully explored and delineated, our evidence shows it is incorrect to claim that grid cells drop out of deep networks when simply trained to path integrate.} 

If path integration is an insufficient task for grid cells to emerge, what might be sufficient? Prior work showing that grid cells are a powerful neural code with exponential capacity and error correction abilities \cite{fiete_what_2008, sreenivasan_grid_2011, mathis_optimal_2012, wei2015principle} suggests a potential direction: our hypothesis is that the grid code emerges by directly optimizing the learned representation for space, in a coding-theoretic sense, so that coding states are well-distributed within the available coding volume. Work in this direction exists \cite{gao2019learning,gao2021path,dorrell2022actionable,xu2022conformal}, and a recent work of ours, ``Self-Supervised Learning of Representations for Space Generates Multi-Modular Grid Cells" \cite{schaeffer2023self}, demonstrates that training deep neural networks with self-supervised learning to both path integrate and optimize the packing of the code in the coding space can generate multi-modular grid cells - without any supervised targets.

\section{Acknowledgements}

IRF is supported by the Simons Foundation through the Simons Collaboration on the Global Brain, the ONR, the NSF, and the K. Lisa Yang ICoN Center. SK acknowledges support from NSF grants No. 2046795, 1934986, 2205329, and NIH 1R01MH116226-01A, NIFA 2020-67021-32799, the Alfred P. Sloan Foundation, and Google Inc. RS is supported by a Stanford Data Science Scholarship. MK is supported in part by a MathWorks Science Fellowship and the MIT Department of Physics.

\clearpage

\bibliography{references,mikail}

\clearpage

\appendix

\section{Further Investigation of Ganguli et al.'s Gaussian Result}
\label{app:sec:sorscher_2022_rebuttal_gaussian}

\begin{figure}[b]
    \centering
    \includegraphics[width=0.5\textwidth]{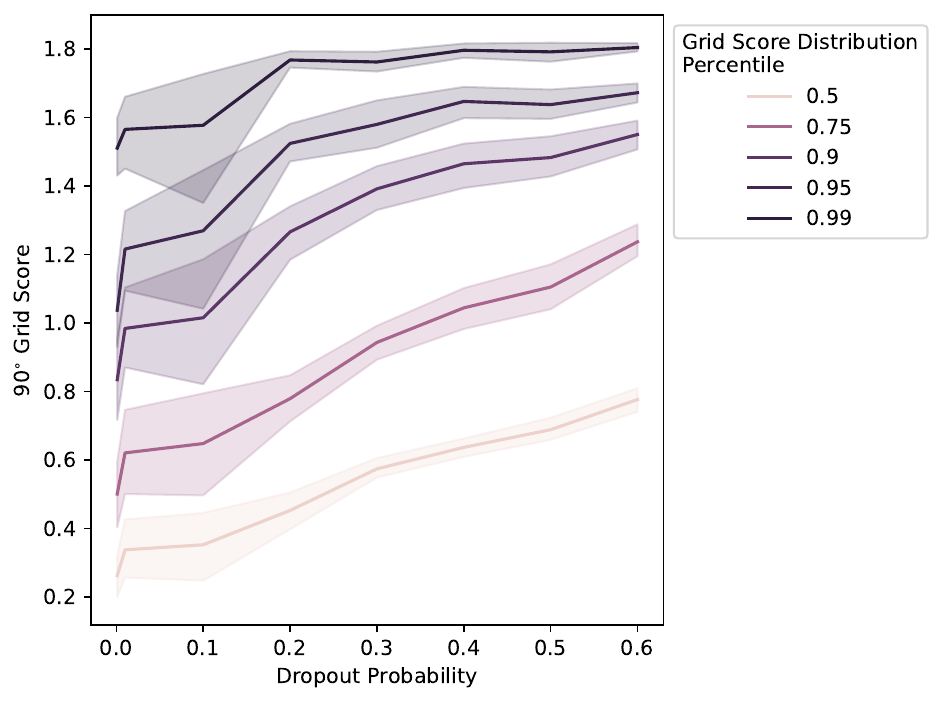}
    \includegraphics[width=0.45\textwidth]{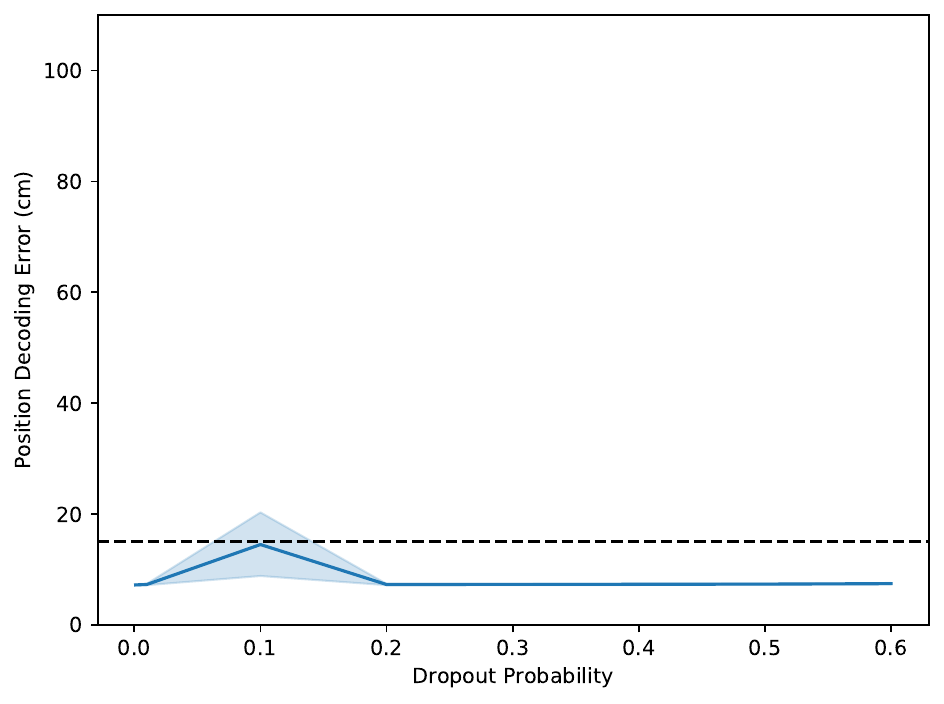}
    \caption{Left: High dropout probability increases the $90^{\circ}$ grid score distribution. Right: High droout does not increase path integration performance.}
    \label{app:fig:sorscher_2022_rebuttal_dropout}
\end{figure}

\begin{figure}[b]
    \centering
    \includegraphics[width=0.5\textwidth]{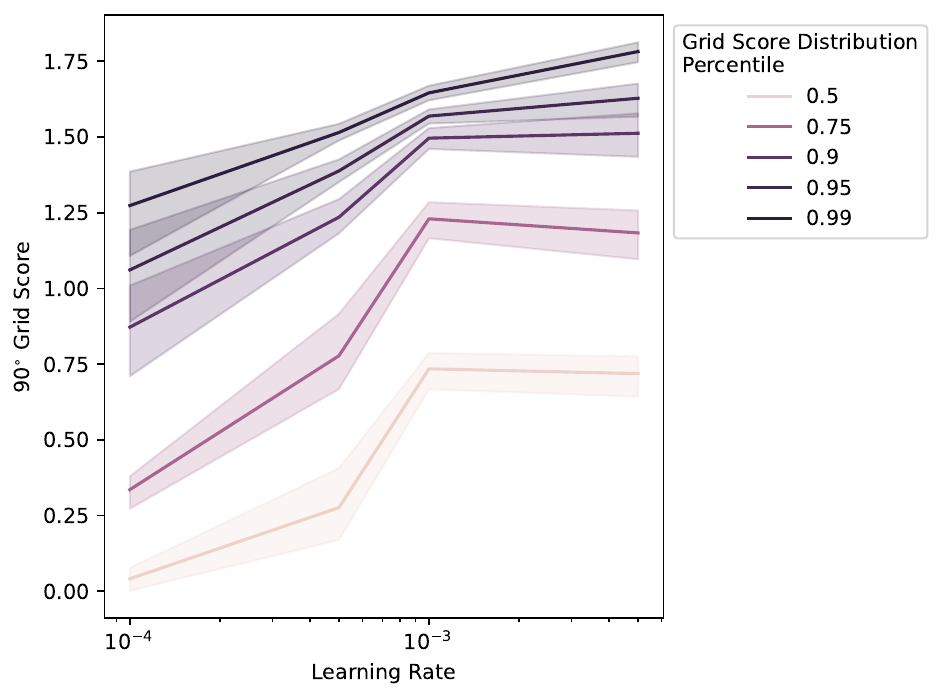}
    \includegraphics[width=0.45\textwidth]{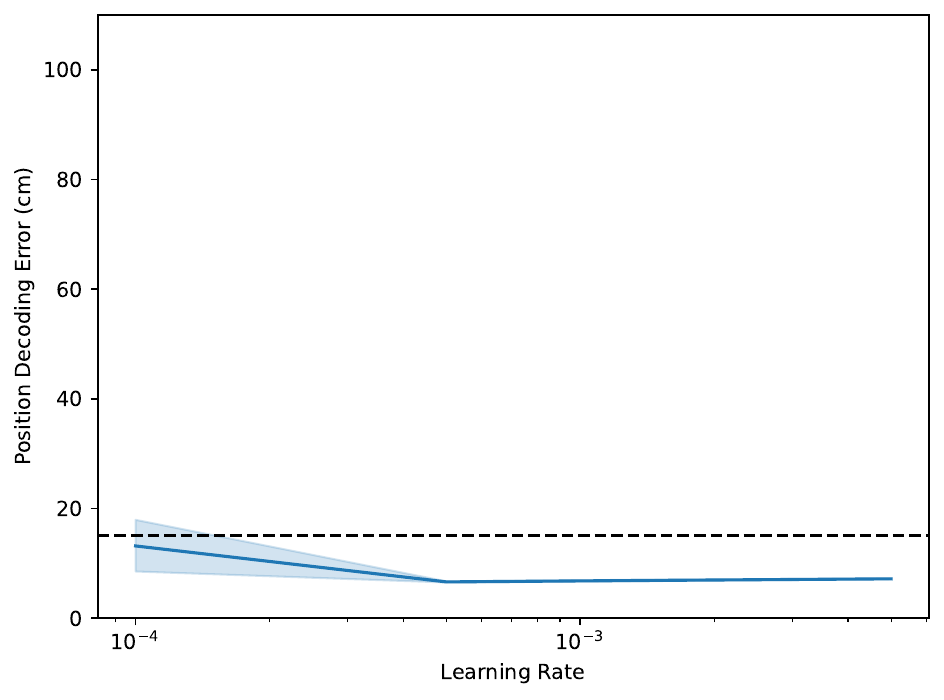}
    \caption{Left: High learning rate increases the $90^{\circ}$ grid score distribution. Right: High learning rate does increase path integration performance.}
    \label{app:fig:sorscher_2022_rebuttal_learning_rate}
\end{figure}

\begin{figure}
    \centering
    \includegraphics[width=0.5\textwidth]{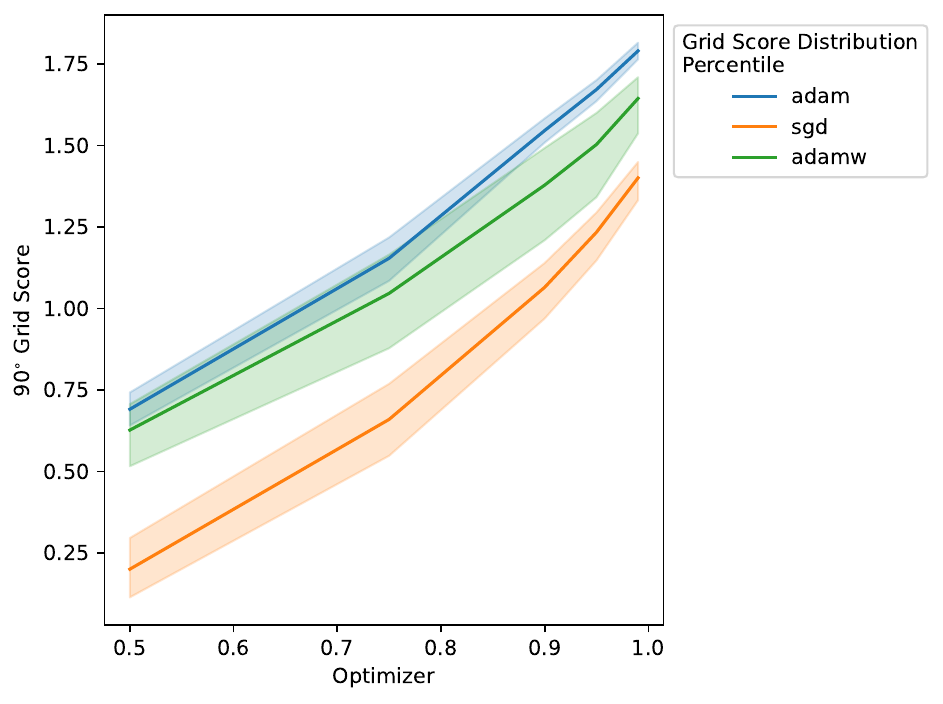}
    \includegraphics[width=0.45\textwidth]{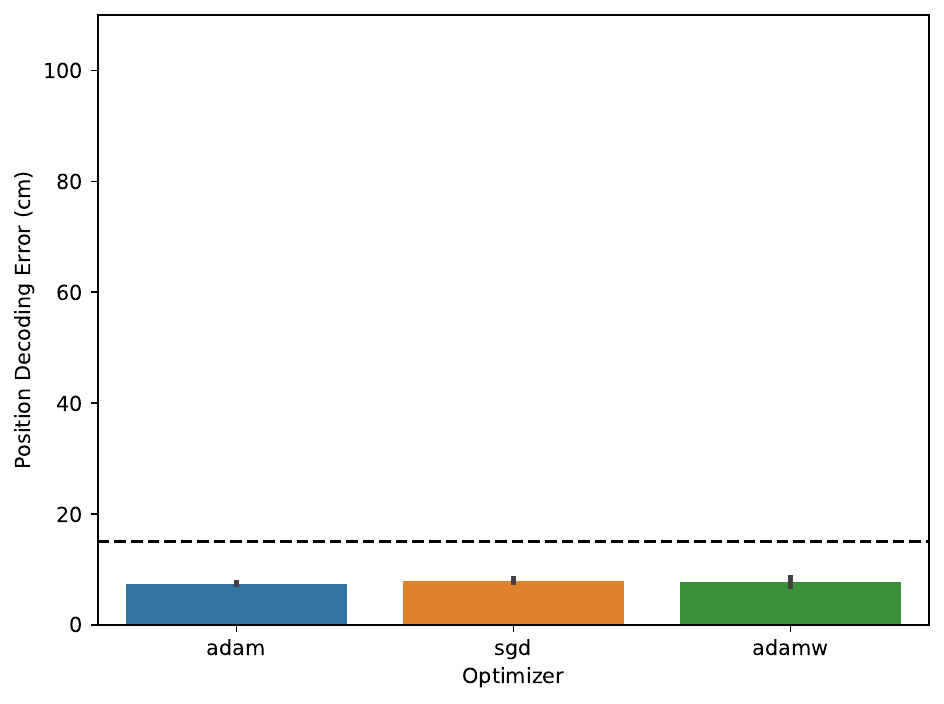}
    \caption{Left: Specific choice of optimizer increases the $90^{\circ}$ grid score distribution. Right: Specific choice of optimizer does not increase path integration performance.}
    \label{app:fig:sorscher_2022_rebuttal_optimizer}
\end{figure}

We include additional results from investigating the newly released code of Ganguli et al. 2022 \cite{sorscher_when_2022} that claims to produce square lattices from Gaussian supervised targets. We find that a combination of additional implementation details unrelated to path integration performance are critical for lattice-like tuning. Specifically, (1) high dropout probability e.g., 50\% (Fig. \ref{app:fig:sorscher_2022_rebuttal_dropout}), (2) high learning rate, e.g., 0.1 (Fig. \ref{app:fig:sorscher_2022_rebuttal_learning_rate}) and (3) choice of optimizer, e.g., Adam (Fig. \ref{app:fig:sorscher_2022_rebuttal_optimizer}) increases the $90^{\circ}$ grid score distribution without improving path integration performance. 

\clearpage

\section{Reproduction \& commentary on the Unified Theory of \cite{sorscher_unied_2019}}
\label{app:theory}

Here, we reproduce the theory of \cite{sorscher_unied_2019,sorscher_unified_2020}, highlighting the simplifying assumptions made by the authors, to provide an explanation for the mismatch between theory and simulations of trained path integrating networks. Further, we reproduce, very briefly, the theory of pattern formation in first-principles continuous attractor models, highlighting similarities and differences between these two theories.

This theory does not deal with dynamics of path integration or learning dynamics of a deep recurrent network from initialization, but rather concerns the problem of readout reconstruction. This leads us to the first assumption:

\textbf{Assumption 1 (A1)}: The hypothetical hidden representation $G \in \mathbb{R}^{n_x \times n_g}$ is some function of space. Here $n_x$ is the number of spatial locations and $n_g$ is the number of hidden units. This is a subtle but significant assumption because, for recurrent networks given velocity inputs, the networks' representations are not a function of space, but rather develop into a function of space (i.e., builds a continuous attractor) over the course of training. For a better understanding of why the assumption of building a continuous manifold of fixed points is significant, see literature of the theory of continuous attractors, which is briefly reviewed in \cite{khona_attractors_2022}.

Under A1, consider a feedforward mapping $\hat{P} \defeq G W$ where $W \in \mathbb{R}^{n_g \times n_p}$. Here $n_p$ is the number of readout units. One can define the readout reconstruction error as the mean square loss between the readout target $P \in \mathbb{R}^{n_x \times n_p}$ and prediction $\hat{P} \defeq G W$ :
\begin{equation}
    \mathcal{E}(G, W) \defeq ||P - \hat{P}||_F^2 = ||P - G W ||_F^2
\end{equation}

\textbf{Assumption 2 (A2)}: Linear readout $W$ relaxes, reaching its optimum much faster than $G$ changes, so that we can replace $W$ with its optimal ordinary least squares value for fixed $G, P$:
\begin{equation}
    W^*(G, P) = (G^T G)^{-1} G^T P
\end{equation}

Substituting $W^*(G,P)$ into the loss for $W$ yields the error as a function of $P$ and $G$:
\begin{equation}
    \mathcal{E}(G, P) = ||P - G (G^T G)^{-1} G^T P||_F^2
\end{equation}

\textbf{Assumption 3 (A3)}: $G$'s columns can be made orthonormal i.e. $G^T G = I_{n_g}$.This means that each grid unit has the same average firing rate over all space, and that any 2 grid cells do not overlap. There is no provided reason why the dynamics of (stochastic) gradient descent in RNNs, with or without regularization, can or should find this solution.

Then, we can write down a Lagrangian for this optimization problem, using a Lagrange multiplier $\lambda$, for this constraint:
\begin{equation}
\mathcal{L} = -\mathcal{E}(G,P) - \lambda (G^T G - I_{n_g})
\end{equation}

We can then set $G^T G$ to $I$ in the error term and the Lagrangian is now written as:
\begin{align}
    \mathcal{L} &= - ||P - G G^T P||_F^2 - \lambda (G^T G - I_{n_g}) \\
    \mathcal{L} &= - \text{Tr}[(P- G G^T P)^T(P- G G^T P)] - \lambda (G^T G - I_{n_g})
\end{align}
Here, the identity $||M||_F ^2 = \text{Tr}(M^TM)$ has been employed.\\
The trace term can be simplified further using the cyclic permutation property of Trace: $\text{Tr}(ABC) = \text{Tr}(CAB)$, 
\begin{align*}
    \text{Tr}[(P- G G^T P)^T(P- G G^T P)] &= \text{Tr}(P^TP) + \text{Tr}[(G G^T P)^T(G G^T P)] - \text{Tr}(P^TGG^TP) - \text{Tr}(P^TGG^TP)\\
    & = \text{Tr}(P^TP) + \text{Tr}[G^T(PP^T)G(G^TG)] - 2\text{Tr}(G^TPP^TG)
\end{align*}
Here $\Sigma \defeq \frac{1}{n_p} P P^T \in \mathbb{R}^{n_x \times n_x}$ is the readout spatial correlation matrix. We can also drop the $G$ independent term above. Using the trace identity again, this term simplifies to $\text{Tr}(G^T PP^T G)$. Hence the total simplified Lagrangian is then:

\begin{align}
    \mathcal{L} &= \text{Tr} \big[G^T \Sigma G - \lambda (G^T G - I_{n_g}) \big]
\end{align}

Considering gradient learning dynamics, one gets the following evolution equation for $G$:
\begin{equation}
    \frac{d}{dt}G = \nabla_G \mathcal{L} \Rightarrow \frac{d}{dt}G = \Sigma G - \lambda G
\end{equation}

\cite{sorscher_unied_2019,sorscher_unified_2020} then simplify further analysis by considering a single grid unit. This corresponds to replacing the $n_x \times n_g$ matrix $G$ by the $n_x \times 1 $ column vector $g$:
\begin{equation}\label{eq:gradientdynamics}
    \frac{d}{dt}g = \Sigma g - \lambda g
\end{equation}

Thus, this linear dynamical system captures how the pattern $g$ of a unit evolves with gradient learning. Further, \cite{sorscher_unied_2019,sorscher_unified_2020} conclude the eigenvectors corresponding to the subspace of the \textit{top eigenvalue} form the optimal pattern, since these eigenvectors will grow exponentially with the fastest rate.

\textbf{Assumption 4 (A4)}: The readout spatial correlation $\Sigma$ is translation-invariant over space i.e. $\Sigma_{x, x'} = \frac{1}{n_p} \sum_{i=1}^{n_p} p_i(x) p_i(x') = \frac{1}{n_p} \sum_{i=1}^{n_p} p_i(x + \Delta) p_i(x' + \Delta) = \Sigma_{x+\Delta, x'+\Delta} \forall \Delta$. This is a strong assumption on the readouts and not necessarily true in biology \cite{schaeffer_testing_2023}.

\textbf{Assumption 5 (A5)}: The environment has periodic boundaries (or no boundaries, which corresponds to a continuum limit). [An alternative assumption in other parts of the derivations and numerics is {\bf (A5')}: the assumption involves periodic boundary conditions with a small box size $L$; the smallness of the box is important for the prediction of grids from Gaussian units. There are other biological plausibility concerns with making a specific choice for box size to explain grid cell responses, which exhibit high invariance to box size in familiar environments (see paragraph below, on ``The nature of discretization effects'').]
% XXX check edits made above. 
%Checked, they are good.

Under A4 and A5, the eigenmodes of $\Sigma$ are exactly Fourier modes across space and thus form a periodic basis. The normalized eigenvectors are indexed by their wavelength, $\mathbf{k}$ and are denoted $f_{\mathbf{k}}$ with corresponding eigenvalue $\lambda_{\mathbf{k}}$.

To calculate this eigenvalue, \cite{sorscher_unied_2019,sorscher_unified_2020} use Fourier theory:
\begin{align*}
    \Sigma f_{\mathbf{k}} &= \lambda_{\mathbf{k}} f_{\mathbf{k}}\\
    \implies \lambda_{\mathbf{k}} &= f_{\mathbf{k}}^{\dagger}\Sigma f_{\mathbf{k}}
\end{align*}
Here $f_{\mathbf{k}}^{\dagger}$ denotes the conjugate of the eigenvector $f_{\mathbf{k}}$. 

Next, they rewrite in component form: $\Sigma_{x,x'} = 1/n_p (PP^T)_{x,x'} = 1/n_p \sum_{i=1}^{n_p}p_i(x)p_i(x')$
\begin{align*}
     \lambda_{\mathbf{k}} = f_{\mathbf{k}}^{\dagger}\Sigma f_{\mathbf{k}} &= \sum_{i=1}^{n_p}\sum_{x,x'}\dfrac{1}{n_p}f^*_{\mathbf{k}}(x')p_i(x)p_i(x')f_{\mathbf{k}}(x)\\
     &= \dfrac{1}{n_p} \sum_{i=1}^{n_p} \left(\sum_x p_i(x)f^*_{\mathbf{k}}(x)\right) \left(\sum_{x'}p_i(x')f_{\mathbf{k}}(x') \right)\\
     &= \dfrac{1}{n_p}\sum_{i=1}^{n_p}\tilde{p}^*(k) \tilde{p}(k) = |\tilde{p}(k)|^2\\
     \implies \lambda_{\mathbf{k}} &= |\tilde{p}(k)|^2
\end{align*}

So, \cite{sorscher_unied_2019,sorscher_unified_2020} conclude that the eigenvalue corresponding to eigenvectors with wavelength $k$ is given by the corresponding power of the Fourier spectrum of the readout correlation matrix $\Sigma$. And thus the optimal pattern is the one which has the highest Fourier power in $\Sigma$.

Further, \cite{sorscher_unied_2019,sorscher_unified_2020} consider the effect of non-negativity perturbatively in the readout regression framework 
% XXX but how neural? they're adding a nonlinearity to the linear dynamics of g in Eq. 9; and this is after they have treated the problem of G-> P transformation as a linear regression problem. 
%Yes, it is not neural, I have changed the language to reflect this.
by phenomenologically adding a term to the Lagrangian, $ = \int_x\sigma(g)dx$.
\begin{equation}
    \mathcal{L} = \text{Tr} \big[G^T \Sigma G - \lambda (G^T G - I_{n_g}) \big] + \int_x\sigma(g)dx
\end{equation}
In Fourier space, \cite{sorscher_unied_2019,sorscher_unified_2020} show perturbatively that this amounts to a cubic interaction term, which is the leading order term that non-trivially distinguishes between nonlinearities such as ReLU and Sigmoid which break the $g \mapsto -g$ symmmetry and nonlinearities such as Tanh which do not. Again, specializing to the single neuron Lagrangian,
\begin{align*}
    \mathcal{L}_{int} = \int_{k,k',k''} g(k)g(k')g(k'') \delta(k + k' + k'') dk dk' dk''
\end{align*}
This term effectively acts as a penalty for non-negativity. Here, it is important to point out that this cubic term appears not only for non-negativity, but rather \textit{any} function that is not anti-symmetric. Negative activation functions such as slightly shifted Tanh can also have a cubic term. Thus, non-negativity is a special case, as has been noted in \cite{sorscher_unied_2019} Appendix B. We refer to this as \textbf{Assumption 6 (A6)}.\newline
Under this assumption, \cite{sorscher_unied_2019,sorscher_unified_2020} conclude that the optimal pattern consists of a triplet of Fourier waves with equal amplitude and $k-$vectors that lie on an equilateral triangle, at $60^{o}$ to each other.

Next, we examine the Fourier spectrum of a translationally invariant Gaussian readout $f(\Delta x) = \dfrac{1}{\sqrt{2\pi \sigma^2}}\exp(-(\Delta x^2)/2\sigma^2)$ under the assumptions of this theory. For simplicity and to provide intuition we write its Fourier transform in 1d, which is given by another Gaussian. The peak of the Fourier spectrum is at $k=0$, or the DC, non-periodic mode:
\begin{align*}
    \tilde{f}(k) &= \int_{\mathcal{R}}\dfrac{1}{\sqrt{2\pi \sigma^2}}\exp(-(\Delta x^2)/2\sigma^2) e^{ik\Delta x}d\Delta x \\
    &= \exp(-k^2 \sigma^2/2)
\end{align*}

\begin{figure}[h!]
    \centering
    \includegraphics[scale = 0.8]{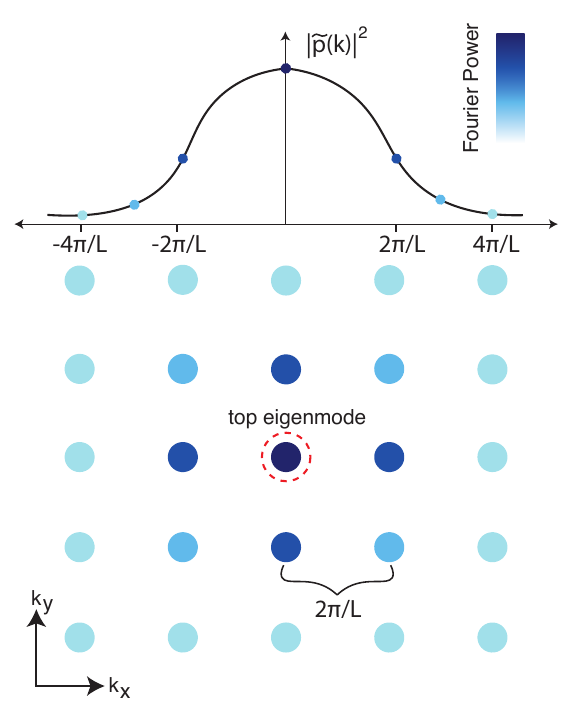}
    \caption{Fourier structure of Gaussian readouts}
    \label{fig:gaussians_no_peak}
\end{figure}

In simulations in a finite environment of length $L$, the allowed frequency modes are discretized with bin-size $2\pi/L$ ({\bf (A5')}. This is shown in Fig.\ref{fig:gaussians_no_peak} as lattice points with the Fourier spectrum overlayed as in \cite{sorscher_unied_2019,sorscher_unified_2020}. Gaussian readouts produce a Fourier spectrum peaked at the central DC mode. This mode has no periodicity and thus the theory for a single hidden unit predicts no lattices in this hidden unit, in the continuum or small-box discrete limit.

Until now, all analysis was performed for a single grid unit. What happens in full multi-cell setting? For this case, \cite{sorscher_unied_2019} shows that the global optimum to the constrained optimization problem: $$\text{max}_G \text{ Tr}(G^T\Sigma G) \text{ such that }G^TG = I_{n_g},$$ i.e. under {\bf (A3)} involves the columns of $G$ spanning the top $n_g$ eigenmodes of $\Sigma$ (Theorem B.2, Appendix B of \cite{sorscher_unied_2019}). Under {\bf (A6)}, \cite{sorscher_unied_2019} also shows that with a Fourier spectrum consisting of a wide annulus (in the discrete setting this corresponds to a Fourier spectrum of rings of different radii), the optimum consists of a hierarchy of hexagonal maps, but only if the Fourier powers of these rings are exactly equal (Lemma B.3, Appendix B). For purely Gaussian spectra (corresponding to Gaussian readouts), the full theoretical solution of this problem depends on specific details of the power spectrum curve (i.e. the width of the Gaussian) since each discrete eigenmode has a different Fourier power which must be taken into account while constructing linear combinations of modes, and thus is not solvable analytically. 
% XXX Above sentence is not at all clear -- what are you saying? Theory is not possible? It is possible but they did numerics instead? 
%An analytical solution is not possible. There is one term coming from the reconstuction part of the loss and one coming from non negativity (which favors hexagons). Ignoring the first (which is possible when all modes have equal fourier power), you can say when hexagons are optimal. When this is not true, the final pattern depends on the specifics of each mode's fourier power. 
%so for this case they simulated the reduced lagrangian and showed that it could create multiple modules -- these come from lattice discretizations.
% XXX Does result only depend on width of Gaussian power spectrum, and if so why hard? And how does it depend on # neurons and L the environment size?
%the width determines the shape of the spectrum. This interacts with L through discretization: example: the second mode will have e^(-1/2) the power of the first if 2pi/L = width of gaussian. This ratio needs to be plugged into the optimization while constructing linear combinations of waves to check which pattern is optimal.
% So a theoretical statement cannot be made that gaussians always lead to a hierarchy of hexagons. Numerically, it seems like there is but there is alot of noise (look at figure 3 of the neurips paper which simulates the lagrangian, there are peaks but there's alot in between too).
Instead, \cite{sorscher_unied_2019} numerically simulates the Lagrangian dynamics with assumption {\bf (A3)} to find a hierarchy of lattices. However, the number of different period lattices that result is related directly to and nearly as numerous as the number of cells. For a large number of neurons, such as the 4096 hidden units used in simulations\cite{sorscher_unified_2022,sorscher_when_2022}, and sufficient wide power spectra, this would mean the optimum solutions would consist of likely hundreds of discrete frequencies (each corresponding to a grid module). This is to be contrasted with the ~4-8 modules estimated to exist in rodents \cite{stensola2012entorhinal}. 

Until NFL, whether the sequence of assumptions in the theory of \cite{sorscher_unied_2019} holds over the course of gradient learning in path-integrating RNNs, and whether their derived global optimal solution is generically gradient-learnable, remained unanswered. Our simulations suggest a negative answer. We also note that in simulations, {\bf (A3)} (the orthonormality of all hidden unit ratemaps), on which this analysis and proof rests, does not seem to hold \cite{sorscher_unified_2022}, since there are multiple cells with the same phase and periodicity. Indeed, in the biological system, grid units are far from orthonormal -- in each grid module, neurons densely (and nearly continuously) tile all phases, meaning that there are always pairs of cells with non-identical but very similar phases and highly overlapping fields. 

%Any units with higher frequencies (i.e. which come from \textit{lower eigenmodes) }observed in simulations of trained path integrators in \cite{schaeffer_no_2022,sorscher_unied_2019,sorscher_unified_2020} and in this paper arise from non-linear effects \textit{not captured} by this theory. Thus this theory fails to account for the structure of ratemaps in this broad class of trained path-integrators.

\clearpage
\paragraph{Theory of pattern formation in continuous attractor networks} Next, we briefly summarize the theory of pattern formation in continuous attractor networks\cite{khona2022smooth,burak2009accurate}. For a full theoretical treatment see \cite{cross1993pattern,ermentrout1998neural}. Neurons are arranged on a sheet with sheet coordinates denoted by $n$. These models assume translation-invariant connectivity between neurons on a neural sheet: $W(n,n') = W(|n-n'|)$, which is specified by a kernel of interaction. For a network of N neurons, the equations describing their dynamics are given by the following system of equations:
\begin{equation}
    \frac{d s(n,t)}{d t} + \frac{s(n,t)}{\tau} = \phi \left[\sum_{n'=1}^N W(n,n') s(n',t) dn'+ B(n) \right]
\end{equation}

We can take the number of neurons, $N \mapsto \infty$, making $n$ a continuous coordinate on the attractor sheet and use neural field equations to describe the dynamics:
\begin{equation}\label{eq:dynamics}
    \frac{\partial s(n,t)}{\partial t} + \frac{s(n,t)}{\tau} = \phi \left[ \int_{-\infty}^{+\infty}W(n,n') s(n',t) dn'+ B(n) \right]
\end{equation}

This connectivity structure sets up an approximate continuum of fixed points. With the correct kernel, the stable states are periodic. To find the periodicity of the formed pattern, we linearize around an unstable uniform state and use Fourier theory.

First, to identify the unstable uniform state:
\begin{equation}
    \dfrac{s_0}{\tau} = \phi \left[ \int_{-\infty}^{+\infty}W(n-n') s_0(n') dn' + B \right].
\end{equation}

\begin{equation}
    s_0 = \frac{\tau B}{1 - \tau \bar{W}}, \text{ for a ReLU nonlinearity,}
\end{equation}
where $\bar W = \int W(n-n') dn'$.

We then consider a perturbative analysis, by examining the evolution of $s(n,t) = s_0 + \epsilon(n,t)$.

We apply our analysis to the early time evolution of this initial condition, such that $\epsilon(n,t) \ll s_0$ is a small growing perturbation. Inserting our form of $s(n,t)$ in Eq. (\ref{eq:dynamics}), we obtain

\begin{equation}\label{eq:epsdynamics}
    \frac{\partial \epsilon(n,t)}{\partial t} + \frac{\epsilon(n,t)}{\tau}  = \phi'(\bar{W}s_0 + B)\int_{-\infty}^{\infty} W(n-n') \epsilon(n',t) dn'
\end{equation}

Next, we posit the ansatz $\epsilon(n',t) = \epsilon e^{ik \cdot n' + \alpha(k) t}$, where $\alpha(k)$ denotes the wavelength-dependent growth rate of this $\epsilon$ perturbation.

\begin{align*}
    \alpha(k) + 1/\tau &= \phi'(\bar{W}s_0 + B)\int_{\infty}^{\infty} W(n-n') e^{-ik\cdot (n-n')} dn' ,\\
            &= \phi'[\bar{W}s_0 + B]\mathcal{F}[W(n-n')], \\
            &= \tilde{W}(k), \text{ for a ReLU non-linearity, assuming } \bar{W}s_0 + B>0,\label{eq:growthrate}
\end{align*}

where $\mathcal{F}[W(n-n')] = \tilde{W}(k)$ is the Fourier transform of the interaction kernel. Hence, the wavelength that will dominate the formed pattern on the attractor sheet is the one with highest growth rate, which is given by:
\begin{equation}\label{eq:kargmax}
    k^* = \text{argmax}_k \tilde{W}(k),
\end{equation}

Typically, in simulations of mechanistic continuous attractor models, a mexican-hat connectivity profile is used\cite{burak2009accurate}. This corresponds to local excitation and broader inhibition in the network:
\begin{equation}\label{eq:ctsMH}
    W(n,n') = W(|n-n'|) = W(\Delta n) = \alpha_E \exp\left( - \frac{(\Delta n)^2}{2\sigma_E}\right) - \alpha_I \exp\left( - \frac{(\Delta n)^2}{2\sigma_I}\right),
\end{equation}
Here, $\sigma_E, \sigma_I$ and $\alpha_E, \alpha_I$ refer to the widths and amplitudes of the excitation and inhibition respectively.

We can use Fourier analysis to calculate the periodicity of the formed pattern:
\begin{align*}
    \tilde{W}(k) = \mathcal{F}[W(n-n')] =  \alpha_E \sigma_E \exp\left(-\frac{\sigma_E^2k^2}{2}\right) - \alpha_I \sigma_I \exp\left(-\frac{\sigma_I^2k^2}{2}\right)
\end{align*}
Then, using Eq. \ref{eq:kargmax}, we find the Fourier mode dominating the pattern is given by:
\begin{equation*}
    [k^*]^2 = \frac{2}{\sigma_E^2 - \sigma_I^2}\log\left(\frac{\alpha_E\sigma_E^3}{\alpha_I\sigma_I^3}\right).
\end{equation*}

\subsection{Connection between the two theories}

To connect these two theories, \cite{sorscher_unied_2019} alternate between a Lagrangian gradient step (Eq.\ref{eq:gradientdynamics}), and a rectification step to impose non-negativity which is equivalent to applying a ReLU non-linearity. Further they assume a total sum constraint on each grid cell's firing to reduce their dynamics to look similar to mechanistic attractor dynamics with a ReLU non-linearity. This is an assumption that we refer to as \textbf{Assumption 7 (A7)}:
\begin{equation}
    \frac{d}{dt}G =  \sigma \left(\Sigma G\right) - \lambda G
\end{equation}
Thus, here the readout correlation matrix plays the role that the interaction matrix plays in mechanistic models.

\paragraph{The nature of discretization effects}Now, we can see similarities between the readout reconstruction loss and continuous attractor networks. In the former case, the Fourier mode with highest power in the readout correlation function tends to dominate the pattern and in the latter, the Fourier mode with highest power in the interaction kernel determines the pattern. 

An important difference is that in the readout reconstruction setting, the pattern is in \textit{real space}, while in the continuous attractor network the pattern is in the \textit{neural sheet} space. This difference is conceptually critical: It is reasonable to assume that the neural sheet is of a given and fixed size, but it is unreasonable to think of the physical space in which the animal moves as being of a given and fixed size. 

The discretization and finite size lattice effects in the readout reconstruction setting are due to the size (and shape) of the physical room used in readout reconstruction and in trained path integrators. This can lead to artefactual lattice-like patterns in some cells as has been observed in \cite{sorscher_unied_2019,sorscher_unified_2020}. It also predicts that the size and shape of the formed lattice will be affected by the room; this is not true in biology for familiar environments of different sizes \cite{hafting2005microstructure}.

On the other hand, the continuous attractor network is subject to discretization effects at the level of number of neurons $N$ -- this is an invariant constraint on the circuit regardless of the current environment the animal is exploring. The number of neurons in MEC is also large, and thus the continuum limit approximation is a valid assumption.

\end{document}